\documentclass[acmsmall,table,xcdraw]{acmart}
\usepackage{subcaption}
\usepackage{url}
\usepackage{multirow}
\usepackage{hhline}
\usepackage{algorithm}
\usepackage[noend]{algpseudocode}
\usepackage{adjustbox}
\usepackage{sidecap}
\usepackage[utf8]{inputenc}
\usepackage{xcolor}
\usepackage{eso-pic}

\AddToShipoutPictureBG{
  \AtPageLowerLeft{%
    \raisebox{1\baselineskip}{\makebox[\paperwidth]{\begin{minipage}{21cm}\centering
    {\color{gray}This paper has been accepted for publication in ACM Transactions on Internet Technology}
    \end{minipage}}}%
  }
}
\usepackage{wrapfig}
\usepackage{calc}
\usepackage{fancyvrb}

\graphicspath{ {plots/} {figures/} {img/} }

\title{Automated Discovery of Network Cameras in Heterogeneous Web Pages}

\date{\today}

\author{Ryan Dailey}
\email{dailey1@purdue.edu}
\affiliation{%
  \institution{Purdue University}
  \city{West Lafayette}
  \state{Indiana}
  \postcode{47906}
}

\author{Aniesh Chawla}
\email{chawla9@purdue.edu}
\affiliation{%
  \institution{Purdue University}
  \city{West Lafayette}
  \state{Indiana}
  \postcode{47906}
}

\author{Andrew Liu}
\email{liu1846@purdue.edu}
\affiliation{%
  \institution{Purdue University}
  \city{West Lafayette}
  \state{Indiana}
  \postcode{47906}
}

\author{Sripath Mishra}
\email{mishra60@purdue.edu}
\affiliation{%
  \institution{Purdue University}
  \city{West Lafayette}
  \state{Indiana}
  \postcode{47906}
}

\author{Ling Zhang}
\email{lingz2@andrew.cmu.edu}
\affiliation{%
  \institution{Carnegie Mellon University}
  \city{Pittsburgh}
  \state{Pennsylvania}
  \postcode{15289}
}

\author{Josh Majors}
\email{jmajors@purdue.edu}
\affiliation{%
  \institution{Purdue University}
  \city{West Lafayette}
  \state{Indiana}
  \postcode{47906}
}

\author{Yung-Hsiang Lu}
\email{yunglu@purdue.edu}
\affiliation{%
  \institution{Purdue University}
  \city{West Lafayette}
  \state{Indiana}
  \postcode{47906}
}

\author{George K. Thiruvathukal}
\email{gkt@cs.luc.edu}
\affiliation{%
  \institution{Loyola University Chicago}
  \city{Chicago}
  \state{Illinois}
  \postcode{60626}
}

\DefineVerbatimEnvironment{urlverb}{Verbatim}{commandchars=\\\{\}}
\newcounter{urlnum}

\begin{document}

\begin{abstract}
Reduction in the cost of Network Cameras along with a rise in connectivity enables entities all around the world to deploy vast arrays of camera networks. Network cameras offer real-time visual data that can be used for studying traffic patterns, emergency response, security, and other applications. Although many sources of Network Camera data are available, collecting the data remains difficult due to variations in programming interface and website structures. Previous solutions rely on manually parsing the target website, taking many hours to complete. We create a general and automated solution for aggregating Network Camera data spread across thousands of uniquely structured web pages. We analyze heterogeneous web page structures and identify common characteristics among 73 sample Network Camera websites (each website has multiple web pages). These characteristics are then used to build an automated camera discovery module that crawls and aggregates Network Camera data. Our system successfully extracts 57,364 Network Cameras from 237,257 unique web pages. 

\end{abstract}

\renewcommand{\shortauthors}{Dailey et al.}

\maketitle

\section{Introduction}
\label{sec:introduction}

\begin{figure}
    \centering
    \begin{subfigure}{0.33\linewidth}
        \includegraphics[width=\linewidth]{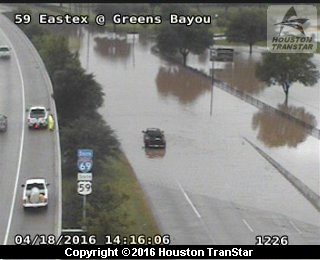}
        \caption{}
        \label{fig:houston_flooding_1}
    \end{subfigure}
    \makebox[0.165\linewidth][c]{}
    \begin{subfigure}{0.33\linewidth}
        \includegraphics[width=\linewidth]{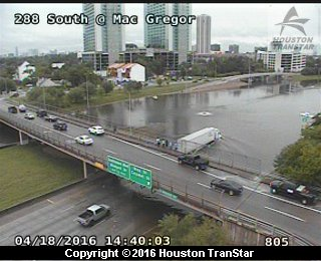}
        \caption{}
        \label{fig:houston_flooding_2}
    \end{subfigure}
    \caption{Real-time Network Camera data gives valuable information during natural disasters such as flooding. These Network Camera images were taken in Houston, Texas from Houston Transtar ~\cite{houston_trans}. (a) shows how first responders could use real-time data to identify individuals that were stranded or needed assistance. (b) shows how Network Camera images could be used to determine a safe path through a flooded city.
    }
    \label{fig:houston_flooding}
\end{figure}

As the cost, ease of use, and Internet bandwidth has improved, more Network Cameras are being deployed by governments, hobbyists, and private entities. Real-time data is data that provides information about the current or near-past. Network cameras returning real-time data offer rich contextual information and are used for weather~\cite{hallowell2005automated}, traffic~\cite{traffic_IPcams}, security~\cite{4525613, 5699107}, and other applications all around the world~\cite{8848161}. These real-time data sources provide distinct advantages over other types of publicly available data because it could be used to study temporally dependent phenomena. For example, Figure~\ref{fig:houston_flooding} shows two instances where real-time Network Camera data could be use to help emergency responders save lives. The images, taken in Houston during a period of intense flooding, show how real-time visual data could be used to identify and rescue individuals stranded in the water (Figure~\ref{fig:houston_flooding_1}). Real-time views of the city could also be used to direct emergency vehicles around heavily flooded city streets (Figure~\ref{fig:houston_flooding_2}). These tasks could not be done with slower methods of visual data collection such as Google Street View~\cite{street_view} where data is rarely collected. 

\begin{figure}
    \centering
    \begin{subfigure}{0.33\linewidth}
        \includegraphics[width=\linewidth]{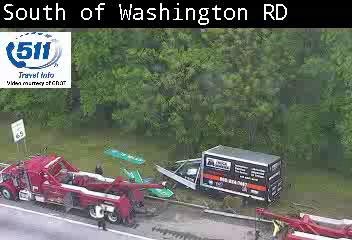}
        \caption{}
        \label{fig:ga_511_1}
    \end{subfigure}
    \makebox[0.165\linewidth][c]{}
    \begin{subfigure}{0.33\linewidth}
        \includegraphics[width=\linewidth]{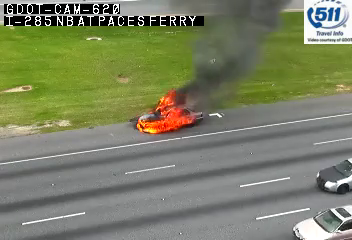}
        \caption{}
        \label{fig:ga_511_2}
    \end{subfigure}
    \caption{Real-time data could be used to identify traffic accidents. These Network Camera images from were taken from traffic cameras installed by the Georgia Department of Transportation ~\cite{ga_511}. (a) shows emergency workers clearing an accident and (b) shows a car fire.}
    \label{fig:ga_511}
\end{figure}

In addition to the rise of real-time Network Camera data, recent developments in deep learning techniques for computer vision have enabled cameras to play a deeper role in threat detection~\cite{millan2012early}, route planning~\cite{Kaseb2015Worldviewandroute894080}, and other applications. Deep learning models require a large amount of training data to achieve high levels of accuracy~\cite{4804817}. Network camera data has been shown to be substantially different from traditional object detection datasets~\cite{dataset_comparison}. Data from Network Cameras often have smaller subjects, more object occlusion, and a greater number of objects per frame compared with traditional datasets like ImageNet~\cite{imagenet}. The real-time data collected from Network Cameras could be used to create new training datasets for computer vision applications. For example, images of crashes and car fires, like those in Figure~\ref{fig:ga_511}, could be used to build computer vision models that alert emergency responders to traffic accidents without human intervention. 

\begin{figure}
    \centering
    \includegraphics[width=0.75\linewidth]{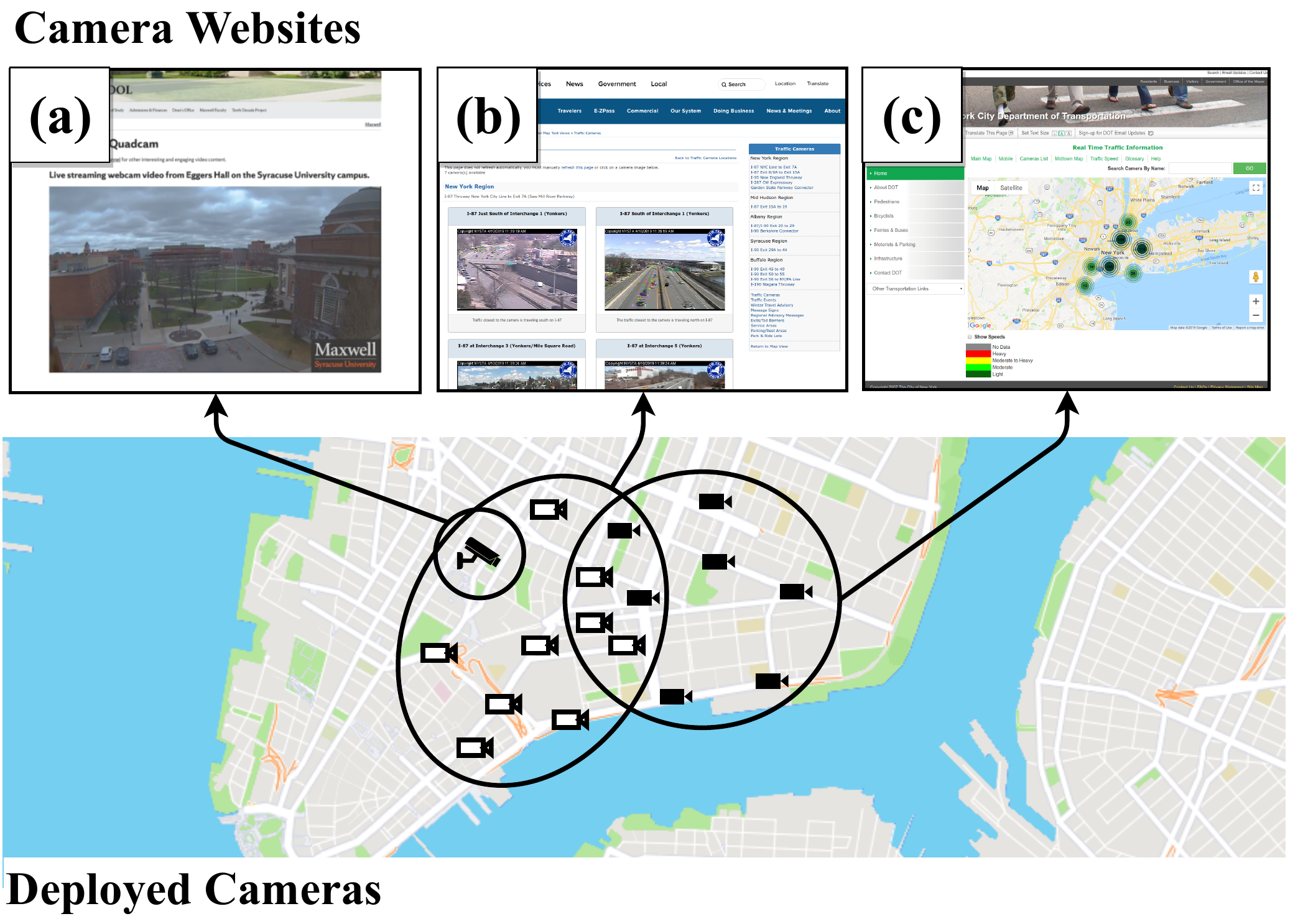}
    \caption{Data from Network Cameras in the same geographical area are distributed across several websites. Each website has a different structure. (a) Syracuse University ~\cite{sys_edu} uses video. (b) New York State Thruway~\cite{ny_state_traffic}, has a list. (c) NYCDOT~\cite{nyc_traffic_cameras}, has a map. None of these websites provide an easy way to retrieve the Network Camera data.}
    \label{fig:camera_dist}
\end{figure}

Despite the ubiquity of real-time public Network Camera data, there is no easy way to collect this data and adopt it to new applications. This is because the data is distributed by thousands of organizations on individual websites across the Internet. For example, universities, regional transportation departments, news stations, and many other organizations in New York State have camera networks deployed. Figure~\ref{fig:camera_dist} shows an example of Network Cameras deployed in New York City. Although the cameras are in overlapping geographical areas, each organization provides its own website to access the camera data. Because cameras providing related data are spread across many heterogeneously structured websites, it is difficult to collect and process the Network Camera data for new applications. To find this data across different websites, one could search for "traffic cameras New York" or "New York webcams" in a search engine and manually determine where the Network Camera data is on the resulting websites. This process is inefficient and does not guarantee that all the Network Camera data for a given area is found. Even if users are able to find a website with relevant Network Camera data, organizations provide no way to download and process the data from their website. Each website has a different user interface and distributes data in different formats (image vs video).

\begin{figure}
    \centering
    \includegraphics[width=1.0\textwidth]{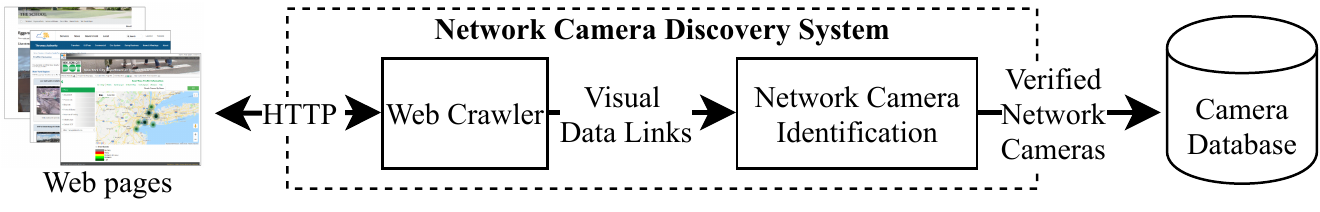}
    \caption{Proposed Network Camera Discovery Process. The system can be broken down into two parts: the Web Crawler module and the Network Camera Identification module. The resulting cameras are then stored in a Camera Database which can be accessed using a RESTful API.}
    \label{fig:system_diagram}
\end{figure}

Our solution for automated Network Camera discovery eliminates the need for human efforts during the discovery process. Our system, shown in Figure~\ref{fig:system_diagram}, creates a database of Network Camera data that users can directly access using a RESTful API. We remove the need to manually search for Network Camera data on the Internet. Using our system, a user can directly query the database API and collect real-time Network Camera data from hundreds of websites. Our Network Camera discovery system consists of two parts: (1) A Web Crawler module that finds data links (2) A Network Camera Identification module that distinguishes between real-time data links and links to static visual data (such as logos). After our system classifies a link to be real-time in nature, the data retrieval information is stored in a database which can be accessed using a public RESTful API. The data can then be used for a variety of purposes including: real-time traffic monitoring, or to build computer vision training datasets to solve new problems. The proposed system allows users to discover Network Cameras through API calls to the database. 

\begin{figure}[h]
    \centering
    \begin{subfigure}{0.33\linewidth}
        \includegraphics[width=\textwidth]{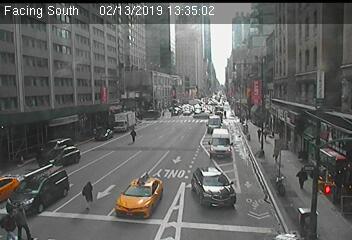}
        \makebox[\textwidth][c]{{\small $t_0$}}
    \end{subfigure}
    \makebox[0.165\linewidth][c]{}
    \begin{subfigure}{0.33\linewidth}
        \includegraphics[width=\textwidth]{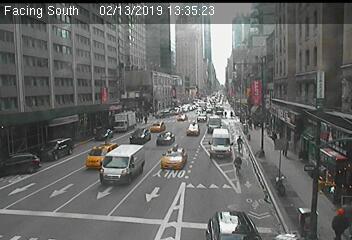}
        \makebox[\textwidth][c]{{\small $t_0+\Delta t$}}
    \end{subfigure}
    \caption{The frames above are from the same Network Camera. These images from the NYC Department of Transportation~\cite{nyc_traffic_cameras} provide real-time data that can be accessed from a single static URL.}
    \label{fig:nyc_traffic}
\end{figure}

\noindent In this work, we define a \textit{Network Camera} as a link with the following properties:
\begin{enumerate}
    \item The link provides visual data, either as a static image or a streaming video.
    \item The data provided by a link is from a statically positioned camera.
    \item The visual data provided by the link is real-time and changes from an initial time ($t_{0}$) to a later time ($t_{0} + \Delta t$). $\Delta t$ can be less than one second to several hours. This feature is highlighted in Figure~\ref{fig:nyc_traffic}.

\end{enumerate}

To develop our solution, we studied a sample of 73 sites that distribute data from Network Cameras (Section~\ref{sec:website_analysis}). We then used this study to identify a set of common data formats that can be used to automatically collect network camera data. We build a system for automated network camera discovery (Section ~\ref{sec:methodology}). We discussed an implementation of our system in Section~\ref{sec:implementation} and evaluated the results in Sections~\ref{sec:results} and ~\ref{sec:analysis}. When our system was tested on the 73 websites, it successfully discovered 57,364 cameras
on the sites. The system checked for Network Camera data from 237,257 unique web pages.
Using the manually labeled data, we evaluate the accuracy of our method. Our camera discovery system achieves 
precision and recall of 98.7\% and 98.2\% respectively on a manually labeled dataset.

\section{Related Work}
\label{sec:related_work}

Many methods have been proposed for creating semantic classifications of multimedia on web pages. For example, Chen et al.~\cite{doi:10.1002/asi.1132} generate descriptions of the content of images using a combination of "high-level" features (file name, hyperlinks, surrounding text) and "low-level" features (image color histogram, shape, and texture). These descriptions can then be searched to enabled users to find useful image data. Similar methods are used by Yuan et al. ~\cite{4734014} to generate a semantic representation of Youtube videos. These works infer properties about the visual data based on contextual information in the web page. Although these works do not identify real-time data, they use the context of the surrounding web page to develop a better understanding of visual data. 

In \cite{Brickley:2019:GDS:3308558.3313685}, Brickley et al. propose a methodology for creating a search engine for publicly available datasets. Datasets are found by parsing RDF (Resource Description Framework), Microdata, and JSON-LD (JavaScript Object Notation for Linked Data) formats. Their broad definition includes visual data but not necessarily real-time data. 

Work has also been done on analyzing Peer-to-Peer streaming networks. In \cite{Vu:2007:MML:1577222.1577227}, Vu et al. proposed using a web crawler to collect information on live Peer-to-Peer TV streams. However, their focus was on network topology and the data they collected was only from one Peer-to-Peer video sharing site PPLive. They do not crawl multiple websites. Furthermore, the data streams on PPLive are TV streams, not streams from Network Cameras. 

The heterogeneity of public sensor data distribution methods was highlighted by Mao et al. in~\cite{Mao:2018:CDC:3274783.3275192}. They propose that public sensor data from governments around the world could be collected and used for other purposes if the data interface was standardized. They suggest that web crawling methods could be used to collect this sensor data. Their focus is on government websites that provide a single web page where public datasets can be downloaded. These datasets however, do not contain real-time image data.

Nath et al.~\cite{nath2006challenges} present a system for visualization of sensor data on a map. Like Mao et al., they note the problem of heterogeneous methods for distributing sensor data in public data repositories. Their solution to this problem is to create an interface for displaying sensor data from various sources in one interface. They do not attempt to automatically highlight sensor data on the internet, instead, their work focuses on how to store and visualize the data after it has been collected. 

The Archive of Many Outdoor Scenes (AMOS) ~\cite{AMOS} project is closely related to our work. They use publicly available Network Cameras to create a dataset of over 330,000,000 images. Jacobs et al. outline the methodology in collecting these cameras as well as analysis done on the images taken over time. To create their dataset, they do not automatically discover the cameras, but rather combine lists provided by Network Camera aggregation sites and manual searches in a search engine. They also allow users to submit their cameras. A comparison of related work can be found in Table~\ref{tab:related_work}.

It is important to acknowledge the privacy implications of this work. In previous work, we discuss the privacy concerns of collecting publicly available multimedia data including Network Camera data~\cite{Lu:2017:PPO:3123266.3133335}. Widen~\cite{4658783} discusses the conditions when privacy can be expected. More specifically, the paper classifies the scenarios based on the subject's location and the observer's location. Recently, the public facial recognition dataset from Microsoft known as MS Celeb~\cite{GuoZHHG16} has been taken down over privacy concerns~\cite{murgia_2019}. 

\begin{figure}
    \begin{subfigure}{0.33\linewidth}
        \centering
        \includegraphics[width=\textwidth]{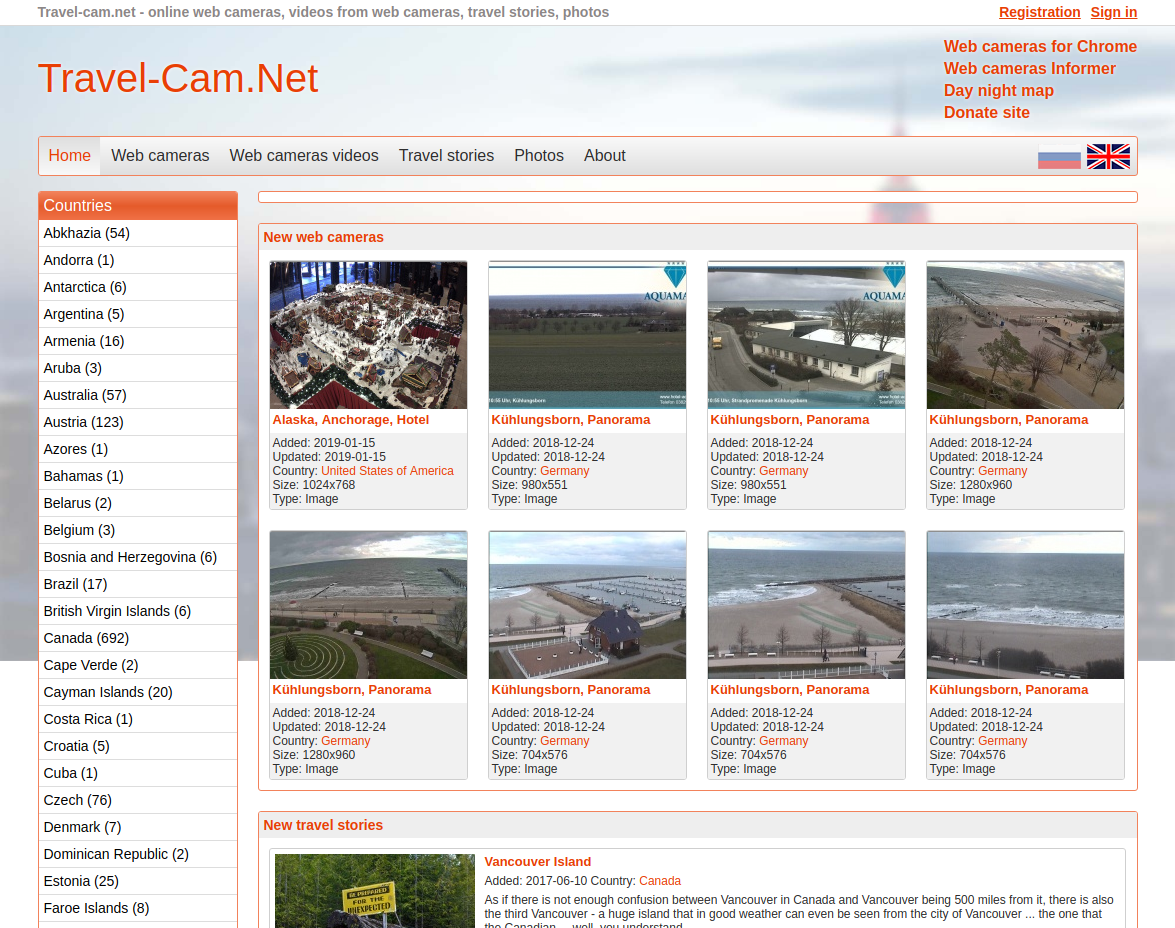}
        \caption{}
        \label{fig:travel-cams}
    \end{subfigure}
    \makebox[0.165\linewidth][c]{}
    \begin{subfigure}{0.33\linewidth}
        \centering
        \includegraphics[width=\textwidth]{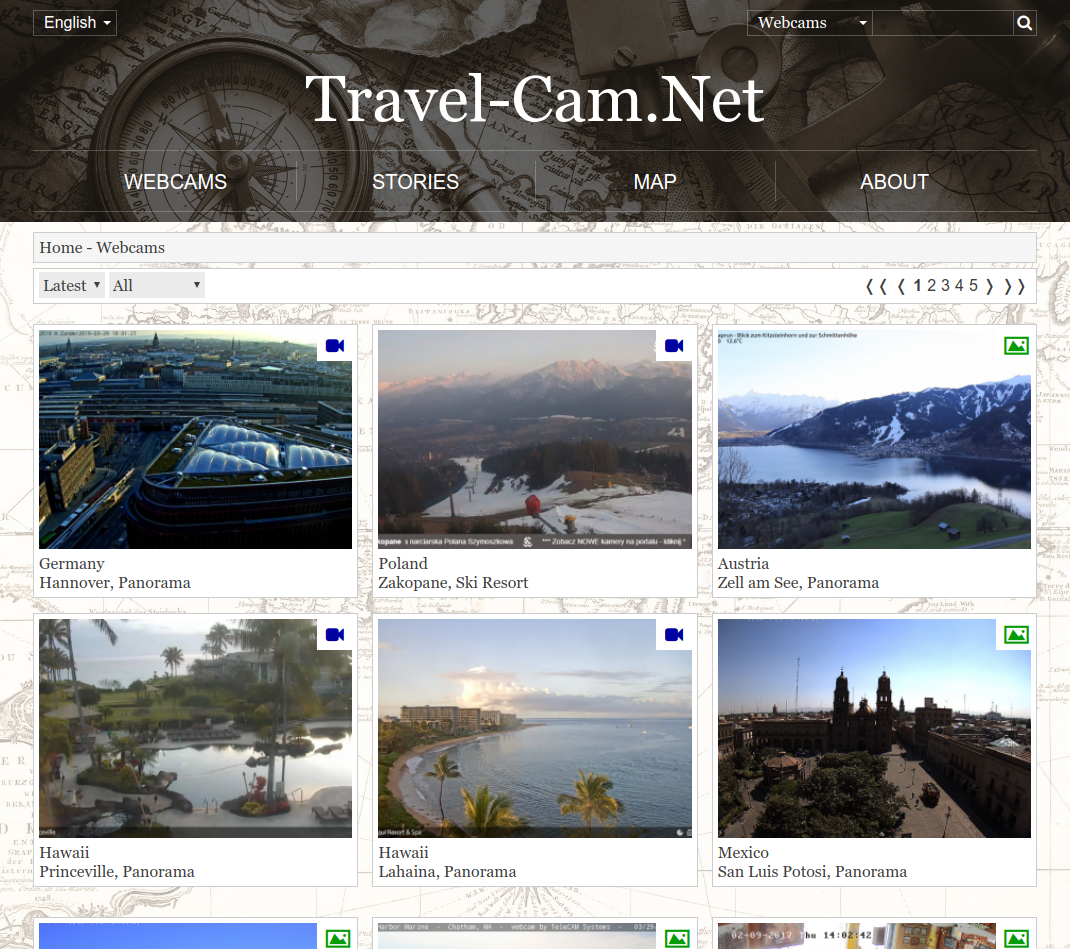}
        \caption{}
        \label{fig:travel-cams_new}
    \end{subfigure}
    \caption{Structure and style of web pages change over time. (a) shows Travel-Cam.net ~\cite{travel-cam} in February 2019. The website has a pull down menu. (b) shows how the site had been redesigned by April 2019. The pull-down menu has been removed.}
    \label{fig:travel_cams_updated}
\end{figure}

Our prior work~\cite{Dailey_imawmw} created a database of Network Camera data using  human-made parsing scripts. In the solution, a parsing script was written for each site individually due to the heterogeneity of website structures. Writing a script for each website makes the camera discovery processes inefficient. Websites change over time and new cameras may be added. For example, Figure~\ref{fig:travel_cams_updated} shows two screen-shots from the Travel-Cam website ~\cite{travel-cam} before and after a redesign. Changing web site structures require manually updating human-made scripts. This paper improves our previous method by creating a general solution that automatically discovers camera data on the Internet. This can lead to easier access to thousands of Network Cameras real-time data.

To our knowledge, this is the study that focuses on automatically collecting real-time video data from heterogeneous websites.

\begin{table}[]
    \centering
    \begin{tabular}{l|c|c|c}
        \multicolumn{1}{c|}{Work} & \begin{tabular}[c]{@{}c@{}}Visual Data \\ Collection\end{tabular} & \begin{tabular}[c]{@{}c@{}}Identify \\ Real-time Data\end{tabular} & \begin{tabular}[c]{@{}c@{}}Automated \\ Data Aggregation\end{tabular} \\ \hline
        Chen et al.~\cite{doi:10.1002/asi.1132} & \checkmark &  & \checkmark \\
        Yuan et al.~\cite{4734014} & \checkmark &  &  \\
        Brickley et al.~\cite{Brickley:2019:GDS:3308558.3313685} &  &  & \checkmark \\
        Vu et al. \cite{Vu:2007:MML:1577222.1577227} &  &  &  \checkmark\\
        Mao et al.~\cite{Mao:2018:CDC:3274783.3275192} &  &  & \checkmark \\
        Jacobs et al.~\cite{AMOS} & \checkmark & \checkmark &  \\
        Dailey et al.~\cite{Dailey_imawmw} & \checkmark & \checkmark &  \\
        \textbf{This Work} & \checkmark & \checkmark & \checkmark
    \end{tabular}
    \caption{This table compares the related work. To our knowledge, this is the only work that combines visual data collection, real-time data identification, and automated data aggregation into one system. }
    \label{tab:related_work}
\end{table}

The major contributions of this paper include
\begin{itemize}
    \item A detailed analysis of the website structures of Network Camera websites.
    \item A general system for automatically aggregating Network Camera data on the Internet. 
    \item A method for determining if a given data link is from a Network Camera providing real-time data.
\end{itemize}

\section{Website Structure Analysis}
\label{sec:website_analysis}
In order to identify Network Camera data from heterogeneous web structures, we create a system as shown in Figure~\ref{fig:system_diagram}. Our system contains a Web Crawler module and a Network Camera Identification module. The Web Crawler module must be able to effectively find and extract Network Camera data links from  web pages. The Network Camera Identification module must be able to differentiate between real-time Network Camera data and other visual data. For these modules to be effective, we must first understand what data formats, website structures, and programming interfaces are most common among Network Camera websites. 

In this section, we identify commonalities that can be used to create a general solution for aggregating Network Camera data. A set of 73 sample websites are analyzed; Table~\ref{tab:site_types} shows the distribution of these sites.  Section~\ref{subsec:camera_formats} explains the data formats, website structures, and programming interfaces of these websites. Section~\ref{subsec:site_org} analyzes how the websites are organized and can be parsed into the database.

\begin{table}
    \centering
    \begin{tabular}{lrrr}
    \textbf{Type} & \textbf{Sites} & \textbf{Estimated Number of Cameras} & \textbf{Cameras/Site}\\ \hline
    Aggregation & 8 & 27909 & 6977 \\
    Traffic & 42 & 21755 & 505 \\
    News Station & 1 & 502 & 502\\
    Tourism & 3 & 5062 & 1687\\
    University & 14 & 31 & 2\\
    Weather & 4 & 2712 & 678\\ \hline
    \textbf{Total} & \textbf{73} & \textbf{57971}
    \end{tabular}
    \caption{Distribution of types of camera websites studied.}
    \label{tab:site_types}
\end{table}

\begin{figure}[t]
    \centering
    \begin{subfigure}{0.20\linewidth}
        \includegraphics[width=\linewidth]{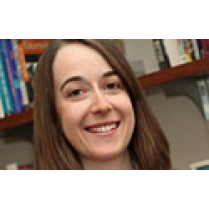}
        \caption{}
        \label{fig:web_assets_b}
    \end{subfigure}
    \makebox[0.04\linewidth][c]{}
    \begin{subfigure}{0.20\linewidth}
        \includegraphics[width=\linewidth]{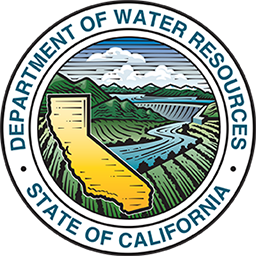}
        \caption{}
        \label{fig:web_assets_a}
    \end{subfigure}
    \makebox[0.04\linewidth][c]{}
    \begin{subfigure}{0.20\linewidth}
        \includegraphics[width=\linewidth]{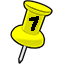}
        \caption{}
        \label{fig:web_assets_c}
    \end{subfigure}
    \makebox[0.04\linewidth][c]{}
    \begin{subfigure}{0.20\linewidth}
        \includegraphics[width=\linewidth]{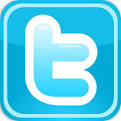}
        \caption{}
        \label{fig:web_assets_d}
    \end{subfigure}
    \caption{Examples of images found on web pages \textit{not} from Network Cameras. (a) Static Photographs~\cite{bu_edu}, (b) Logo, (c) Map markers, (d) Social media logos~\cite{dot_ca_gov}.}
    \label{fig:web_assets}
\end{figure}

\subsection{Data Formats for Network Cameras}
\label{subsec:camera_formats}

Most Network Cameras studied on the sample websites distribute camera data in either static image or streaming video formats. Only one website on the 73 example sites, OnTheSnow ~\cite{onthesnow}, did not use static image or streaming formats. In some cases ~\cite{dot_ca_gov, mo_dot, travel-cam}, sites post a combination of both static images and streaming videos. 
The following paragraphs describe the differences of these data formats.

\subsubsection{Static Images}
Static images such as PNG and JPEG are the most common formats for public Network Camera data. Of the 73 sample websites, 57 provide some static images. These static images are snapshots from the cameras and are updated intermittently. The user must send a new HTTP request to retrieve updated data from the websites.

For most websites, each time an image is updated, the server overwrites the previous image. This means that an HTTP request sent to the same link at $t_0$ and $t_0+\Delta t$ will return different images. Figure~\ref{fig:nyc_traffic} shows two example images downloaded from the same Network Camera data link. Network cameras that have a link format similar to~(\ref{url:baseid}) are the most common of the 57 websites that provide static images. Next we will discuss other link formats that were observed in our sample websites. 

\begin{urlverb}
    <base URL>/<camera id>.jpg \label{url:baseid}
\end{urlverb}

Several sites ~\cite{hb_511_nebraska, rlp_traffic_cameras,utahcommuterlink} have query strings appended to the URL when the user loads the page. In most sites with query strings, the \verb|date-time| field has no effect on the data downloaded from the link. In (\ref{url:queryandtimestamp}) for example, each time the web page is loaded, the \verb|date-time| field is updated. For this group of sites, the query string can be completely removed from the data link and the most recent image data will still be returned. 

\begin{urlverb}
    <base URL>/<camera id>.jpg?<date-time> \label{url:queryandtimestamp}
\end{urlverb}

The above examples show that even within Network Camera websites that distribute data as static images, there is significant heterogeneity to the formatting of the data links. Next, we discuss sites that distribute Network Camera data in streaming video formats.

\subsubsection{Streaming Cameras}

The second most popular format for camera data is streaming formats such as HTTP Live Streaming (HLS), Motion JPEG (MJPG), Real-time Message Protocol (RTMP), and Real-time Streaming Protocol (RTSP). Streaming data is less common, as only 16 sites (22\%) from the sample websites used streaming formats. The most common streaming format is MJPG which makes up about 43\% of streams followed by HLS (38\%) and RTMP (18\%). In some cases, a single site offered more than one streaming format. 

RTMP, RTSP, and HLS formats require embedded video players to view the camera data in browsers. HLS camera data is loaded using an m3u playlist file. In 2 sites, the m3u file is loaded automatically when the video player is loaded and requires no user interaction. For 3 other sites, users must interact with the video player (press a play button) before an HTTP GET request is sent for the m3u file. This type of interaction can make it difficult for cameras to be automatically identified. 

MJPG streams can be directly embedded in a web page and do not need a video player to view. These streams are often from cameras that have HTTP servers inside. These cameras will respond to a variety of HTTP requests allowing the user to get either static image, MJPG video, or HLS video from the camera by sending specific HTTP requests. For some other cameras, the data is distributed using either an HLS stream or an RTMP stream. In some cases ~\cite{511sc, dot_ca_gov}, an RTMP or HLS link can be found in an XHR request (discussed further in Section~\ref{subsec:site_org}). This is usually when the link is loaded in the page automatically and does not require the user to click or interact with the page to load the video data.

\subsection{Network Camera Site Organization}
\label{subsec:site_org}

In this section, we analyze how camera data is organized and presented on the sample websites. The diversity of structures and organizations within sample sites is greater than the formats discussed in Section~\ref{subsec:camera_formats}. The developers of each site use different programming interfaces, making it difficult to find shared characteristics that can be used to create a generalizable aggregation method. Some sites embed camera data links directly in the HTML of the pages, other sites require the user to click a button or scroll on a map to load the camera data.

\begin{figure}
    \centering
  \begin{subfigure}{0.24\linewidth}
        \includegraphics[width=\textwidth]{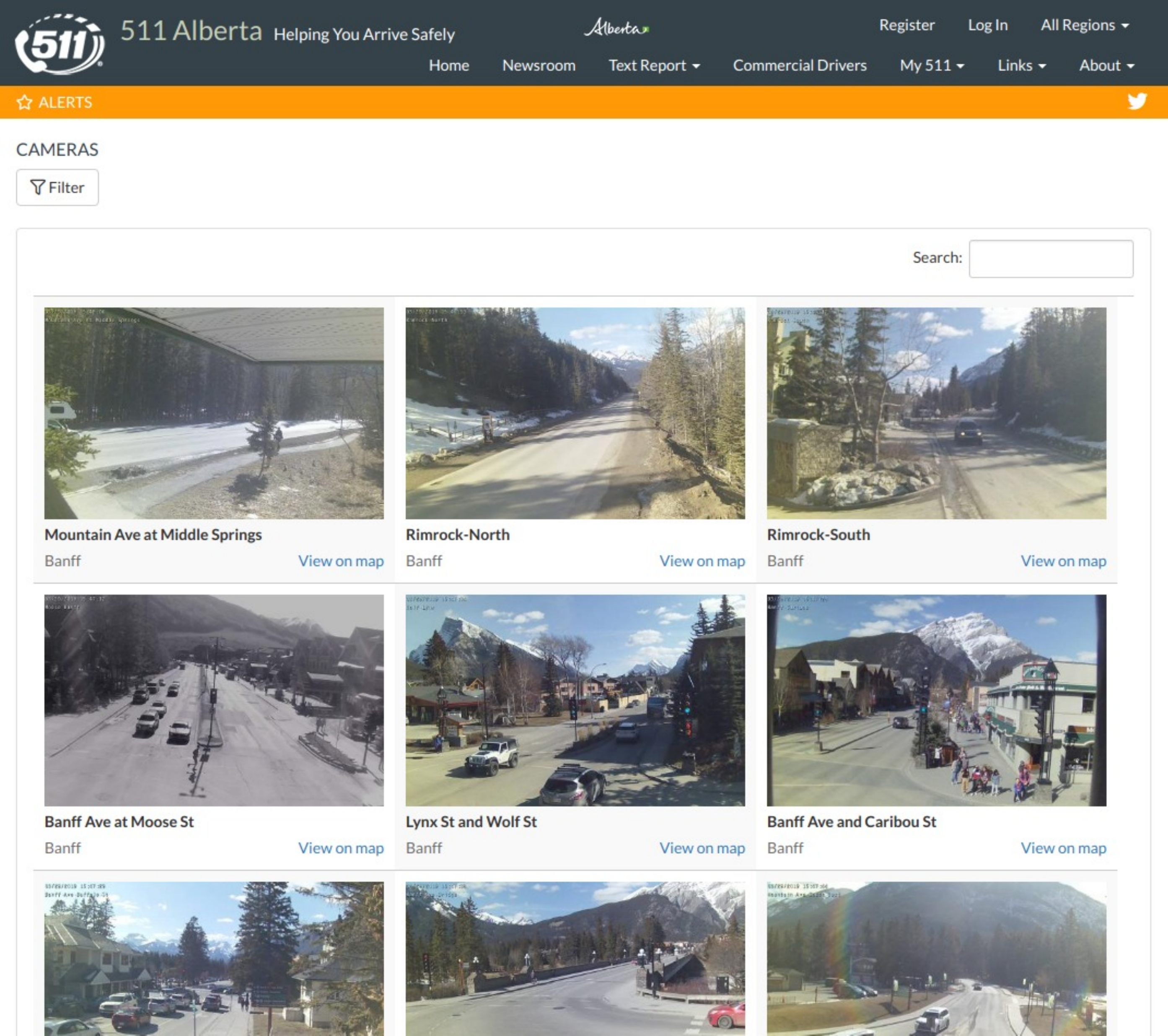}
        \caption{}
        \label{fig:511_alberta_ca_list}
    \end{subfigure}
   \begin{subfigure}{0.24\linewidth}
        \centering
        \includegraphics[width=\textwidth]{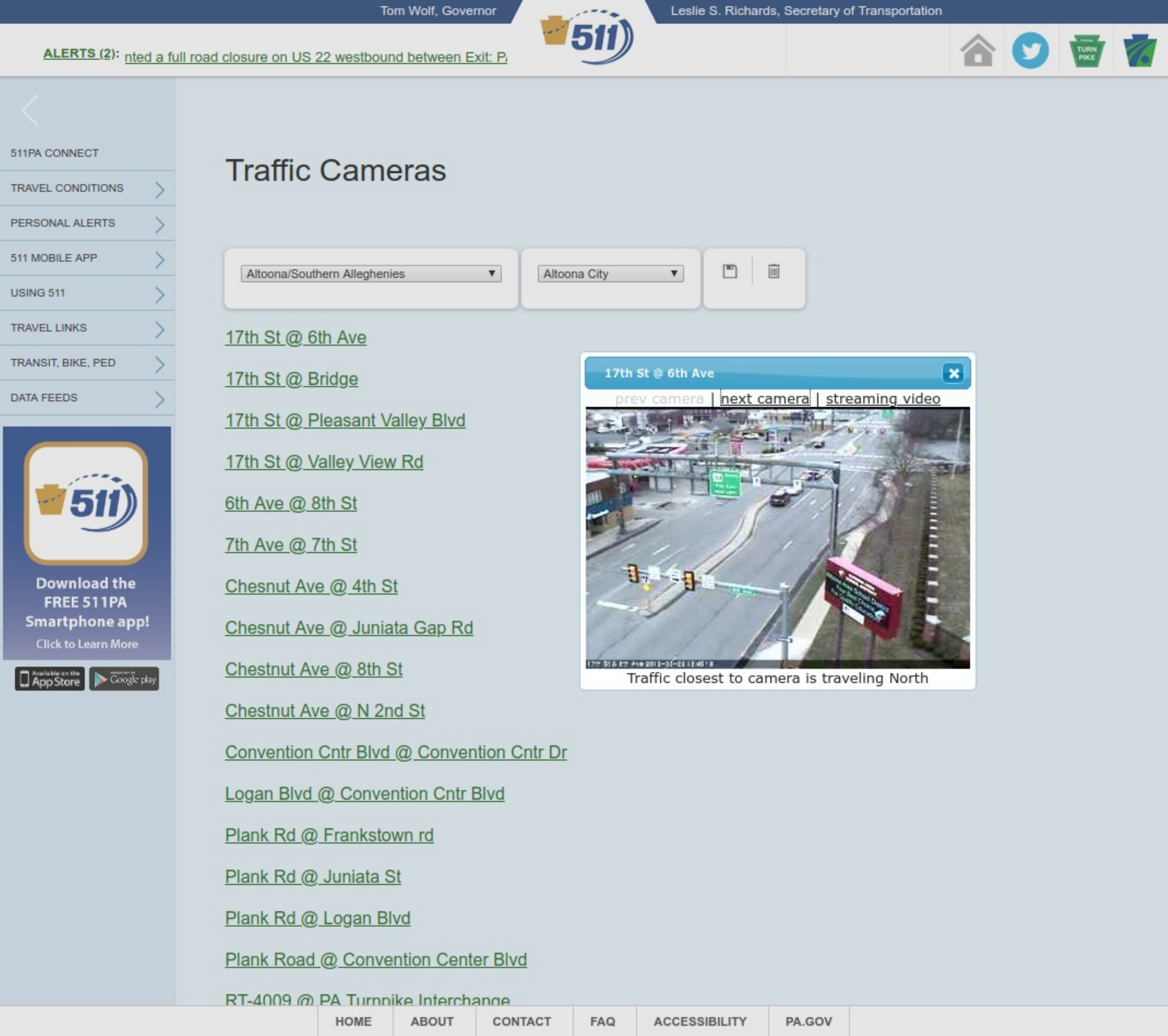}
        \caption{}
        \label{fig:pa_511}
    \end{subfigure}
   \begin{subfigure}{0.24\linewidth}
        \includegraphics[width=\textwidth]{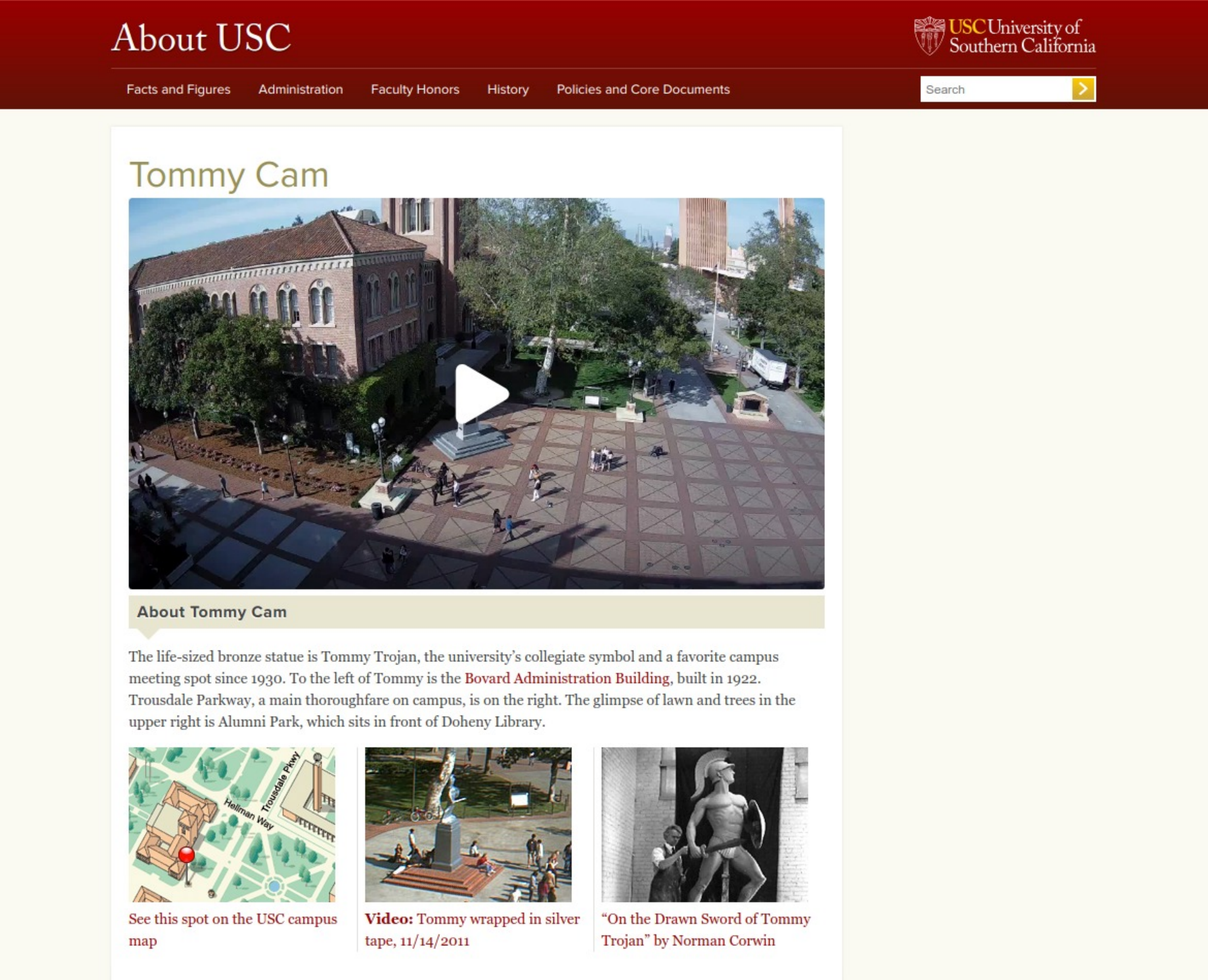}
        \caption{}
        \label{fig:usc_edu}
    \end{subfigure}
	\begin{subfigure}{0.24\linewidth}
		\includegraphics[width=\textwidth]{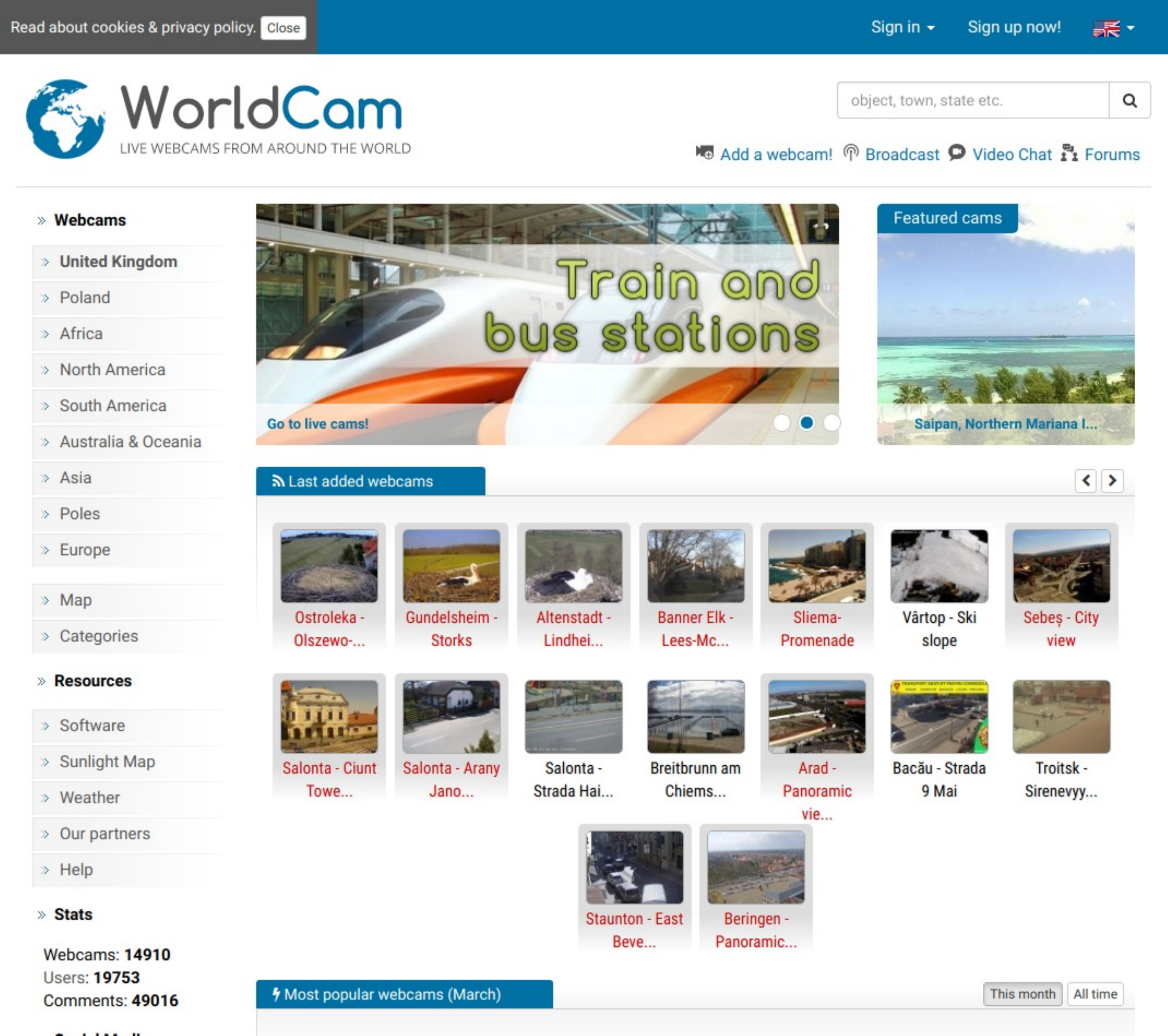}
		\caption{}
		\label{fig:world_cam}
	\end{subfigure}
    \caption{Examples of different list views from Network Camera websites. (a) and (b) are from department of transportation websites ~\cite{511_alberta_ca, pa_511}. (c) shows a Network Camera on a university website  ~\cite{usc_edu}. (d) shows an aggregation site called WorldCam.eu~\cite{worldcam_eu}. This site shows Network Cameras by their geographic locations.}
    \label{fig:list_examples}
\end{figure}

\begin{figure}
    \centering
	\begin{subfigure}{0.24\linewidth}
    	\includegraphics[width=\textwidth]{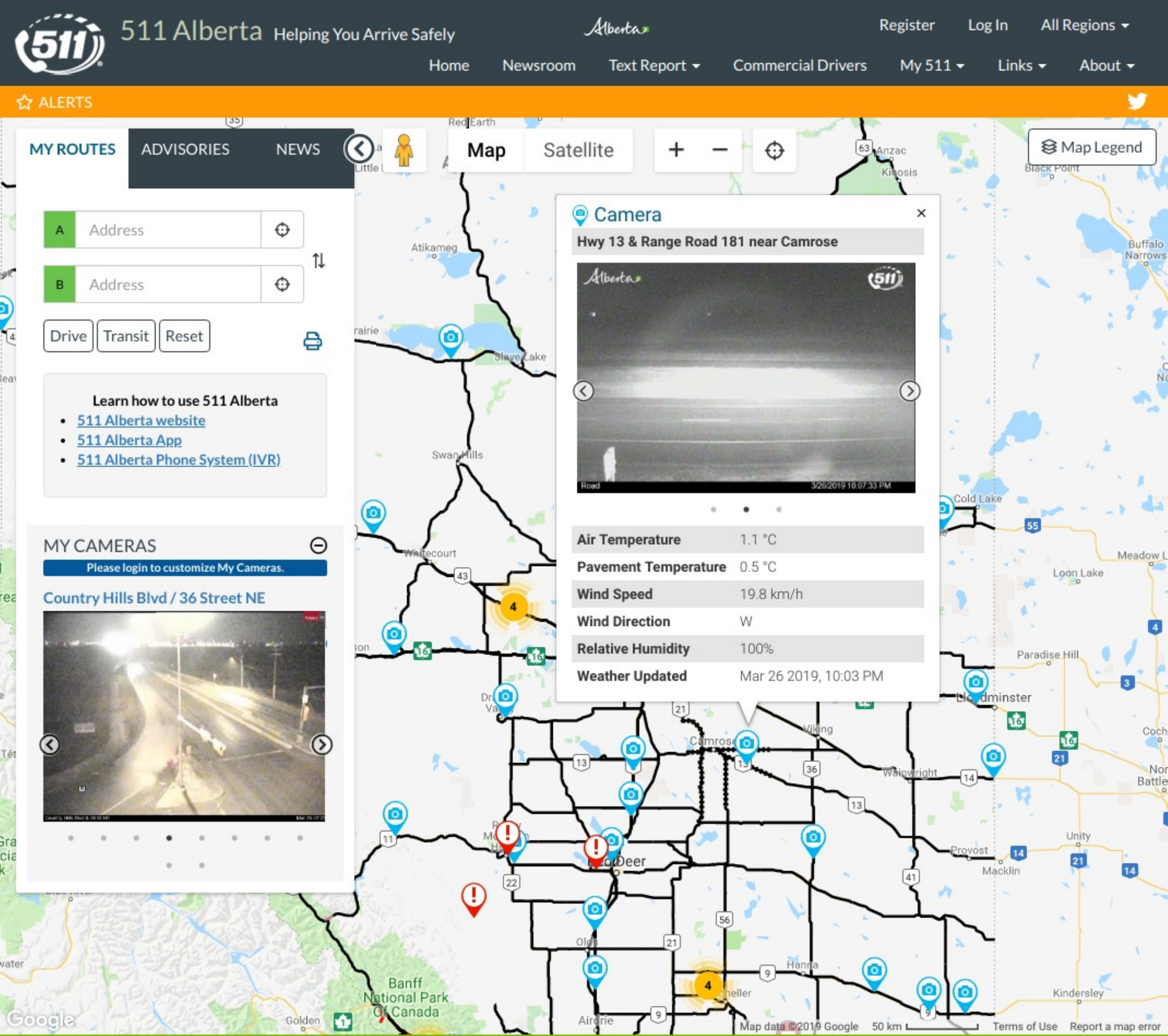}
    	\caption{}
    	\label{fig:511_alberta_ca}
    \end{subfigure}
    \begin{subfigure}{0.24\linewidth}
        \centering
        \includegraphics[width=\textwidth]{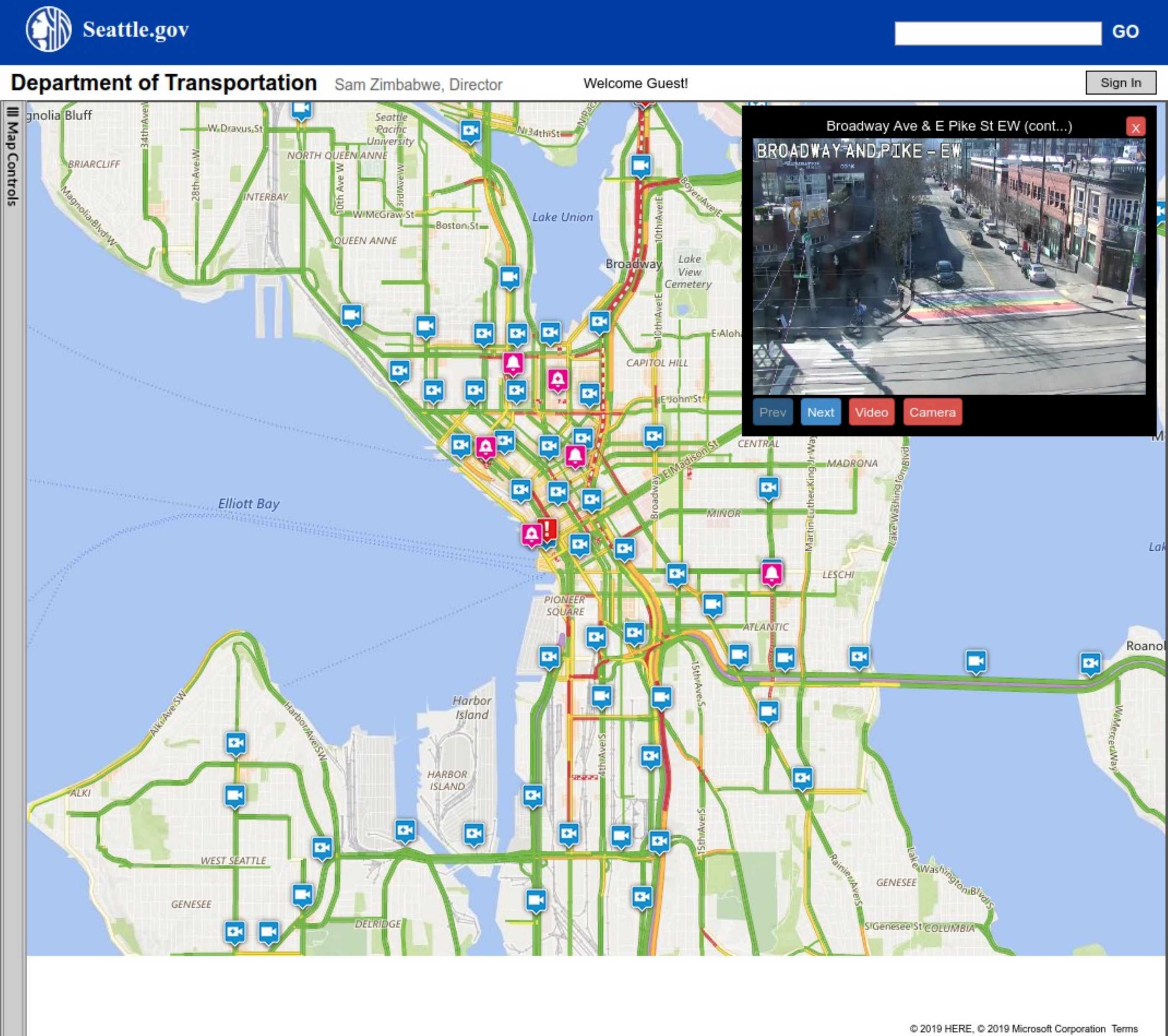}
        \caption{}
        \label{fig:seattle_gov}
    \end{subfigure}
   \begin{subfigure}{0.24\linewidth}
        \includegraphics[width=\textwidth]{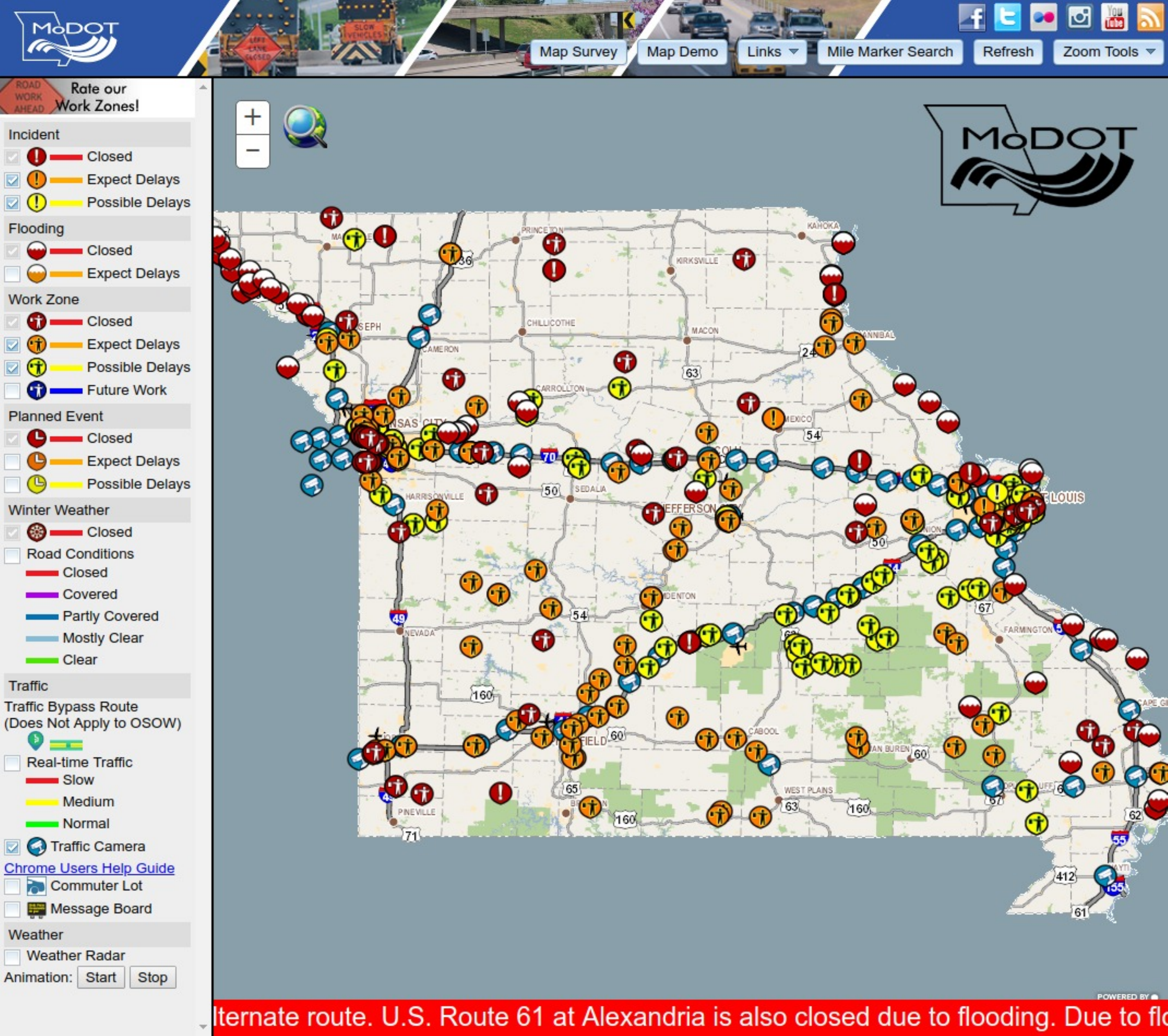}
        \caption{}
        \label{fig:traveler_modot}
    \end{subfigure}
   \begin{subfigure}{0.24\linewidth}
        \includegraphics[width=\textwidth]{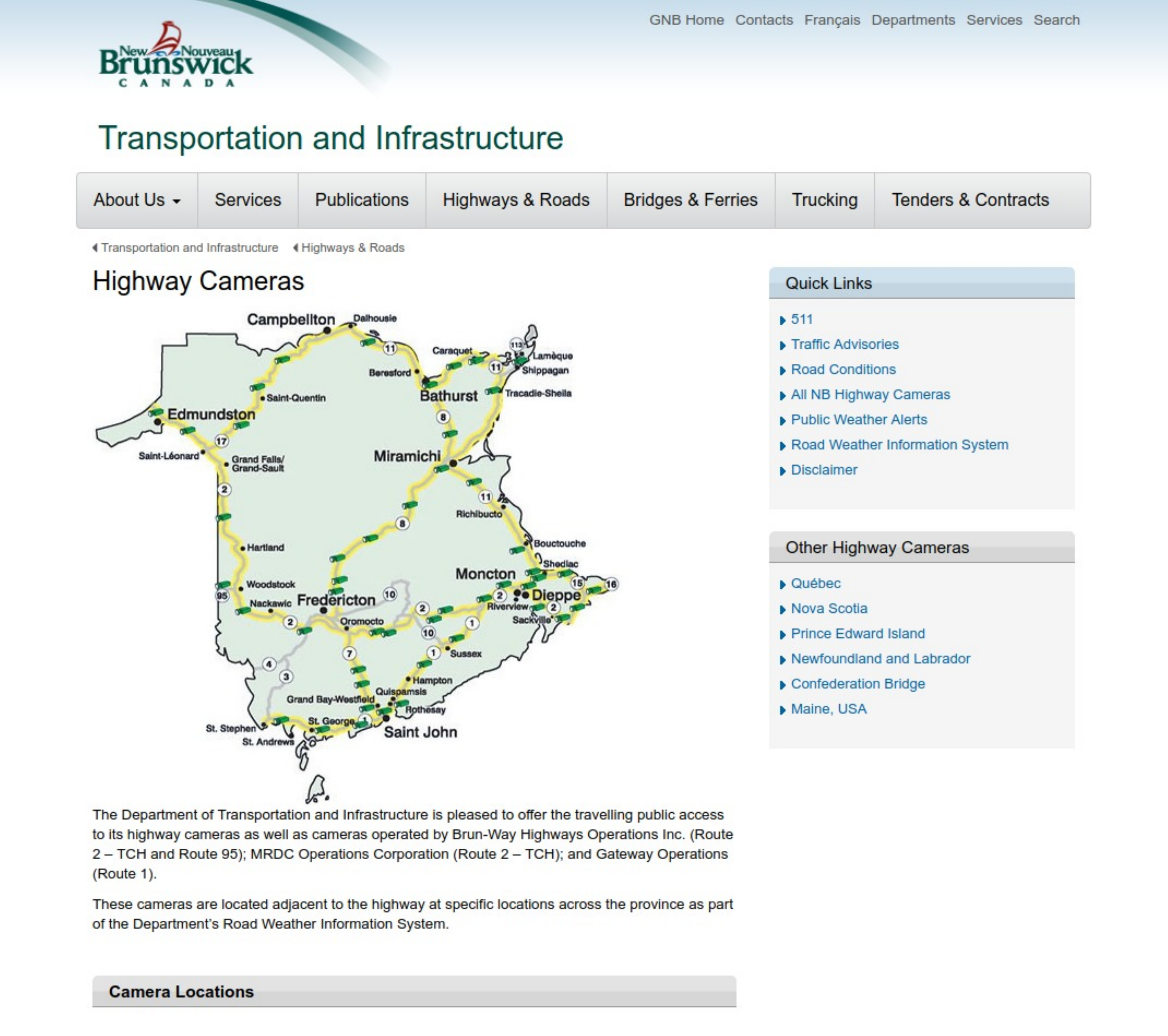}
        \caption{}
        \label{fig:gnb_ca}
    \end{subfigure}
    \caption{Examples of websites that distribute camera data in interactive maps using different APIs. (a) This website displays a popup on Google Maps API~\cite{511_alberta_ca}. (b) The Seattle Department of Transportation uses the HERE Maps API~\cite{seattle_gov}. Clicking on a camera marker will display an image on the top right corner of the map. (c) The Missouri Department of Transportation traffic website~\cite{mo_dot} uses ArcGIS API. (d) When a user clicks on a map marker, another webpage is shown~\cite{gnb_ca}.}
    \label{fig:map_examples}
\end{figure}

Humans can navigate sites to access the camera data but automated solutions may not be able to interact with these views to load the Network Camera data. For example, the traffic website for the City of Tallahassee~\cite{talgov} requires the user to close a pop up menu, select a drop down menu, then select "Cameras" in the drop down menu.
Humans can deduce how to interact with the pages to load the camera data, but these interactions are complicated for machines because there is no standard way to interact with the page.
Figures~\ref{fig:list_examples} and \ref{fig:map_examples} depict the variability within the sample Network Camera sites. 

We identify two methods that are most common for organizing data on the pages: (1) Lists Views (2) Map Views. The following subsections analyze these two methods in more detail. In general, all 73 sample websites can be categorized into methods (1) and (2), but variation exists between implementation details. This means that a method for aggregating data from one website may not work on another website.
In some of the sample sites, both a map and a list are used to organize the data on the same site. For example, Figures~\ref{fig:511_alberta_ca_list} and \ref{fig:511_alberta_ca} show a map and a list both from the Alberta Department of Transportation website~\cite{511_alberta_ca}. 

\subsubsection{List Views}
\label{subsec:list_views}

For 47 of the 73 (64\%) sites provide list views of the Network Camera data. Figure~\ref{fig:list_examples} shows examples of different list views. We can further break down list views by studying how the data links are loaded into the pages. 
In some cases~\cite{kandrive,az511,rlp_traffic_cameras}, real-time image data is embedded in the HTML and loaded directly with the pages, commonly seen in traffic camera websites. Figure~\ref{fig:511_alberta_ca_list} shows a screen-shot from Alberta 511~\cite{511_alberta_ca}. In this example, the site uses an API call to load camera data into the page. The most recent snapshot is then automatically embedded in the HTML of the page. For these sites, simply parsing the HTML and extracting the \verb|<img src="...">| tags yields camera data links. 

In other cases, aggregating the camera data requires user interaction. These sites are more difficult to automatically aggregate due to the range of interactions needed to load the links. In Figure~\ref{fig:pa_511} no image data is loaded into the page until the user selects a region and a route from a drop down menu. After these parameters are selected, the user must select a camera location from the corresponding list. One image can be loaded at a time.

Another type of list view can be observed on aggregation sites~\cite{eisenbahnlivecam,webcamsmania,travel-cam}. These sites rely on user submissions to aggregate large numbers of Network Cameras. These aggregation sites have more cameras than other types of Network Camera websites but in many cases the data is not hosted directly on the site. WorldCam.eu, seen in Figure~\ref{fig:world_cam}, is one such site.

Fourteen websites are from universities. These sites have a large number of pages but few Network Cameras, making it difficult to find cameras on these sites. Often universities have one or two Network Cameras looking at iconic areas around the campus. These cameras are not listed in a table on the site like the previous examples. A Network Camera from University of Southern California~\cite{usc_edu} is shown in Figure~\ref{fig:usc_edu}. This website has only one Network Camera but thousands of web pages.

\begin{table}
    \centering
    \begin{tabular}{lrr}
        \multicolumn{1}{c}{\textbf{API}} & \multicolumn{1}{c}{\textbf{Number of Sites}} & \multicolumn{1}{c}{\textbf{Example}} \\ \hline
        Google Maps & 18 & Figure~\ref{fig:511_alberta_ca} \\
        Openlayers & 8 & \multicolumn{1}{l}{} \\
        ArcGIS & 7 & Figure~\ref{fig:traveler_modot} \\
        HERE & 4 & Figure~\ref{fig:seattle_gov} \\
        Leaflet & 3 &  \\
        HTML & 2 & \multicolumn{1}{l}{Figure~\ref{fig:gnb_ca}} \\
        Other & 1 &  \\ \hline
        \textbf{Total} & \textbf{43} & 
    \end{tabular}
    \caption{APIs used by websites that had interactive maps.}
    \label{tab:map_apis}
\end{table}

\subsubsection{Map Views}
\label{subsec:map_views}

Some sites provide interactive maps. For 43 of the 73 (59\%) sites present camera data in a map. In most cases, the user can zoom and pan around the map to see the geographic locations of the cameras. Figure~\ref{fig:map_examples} shows some examples of websites that use interactive maps. Map interfaces are especially common with traffic camera websites from regional departments of transportation (DoT).

Interactive maps are similar to the list view shown in Figure~\ref{fig:pa_511} because they do not embed the Network Camera data directly in the HTML pages. On these sites, Network Camera images are only shown on the pages when users click on the corresponding map markers. Camera data is displayed in one of three ways: (1) inside a map as shown in Figures~\ref{fig:seattle_gov}, (2) as a pop-up shown in Figure~\ref{fig:511_alberta_ca}, or (3) each map marker links to another web page where the images are displayed~\cite{ib_511ia,511_alberta_ca}.  For most of sample sites with map views, a JavaScript API is used to create the interactive maps. The examples in Figures~\ref{fig:map_examples}abc show maps created with a JavaScript API. Table~\ref{tab:map_apis} lists the different APIs used by the 43 sample sites that have maps. Google Maps~\cite{google_maps} is the most popular API used by the sites accounting for 41\% of all sites that use JavaScript.

To create map markers, JavaScript APIs need geospatial information. In 70\% of cases, the geospatial data comes from an XMLHttpRequest (XHR) to a JSON or XML file. XHR files are found by monitoring the HTTP requests sent by the client during the page load. Eighty-two percent of the geospatial XHR files also contain links to the Network Camera data streams. In many cases, all information that appeared on the map marker when it was clicked was also stored in the the XHR file. Three of the websites contain Network Camera data hard-coded into  JavaScript files instead of a JSON or XML file. 

In 12 sites that use JavaScript API, the geospatial data is generated dynamically and not loaded in XHR files. If the XHR file was found but contained no data links, the link is usually created for each camera using camera IDs from the XHR file. For 5 sites, user interaction is required before the XHR request is sent. For the South African National Roads Agency~\cite{i-traffic_co_za}, as the user zooms and pans the map, XHR requests are sent, loading small sections of the camera data at a time. Zooming and panning operations occur on only this site, but are important to note as they are more complicated for an automated solution to preform. 

Two of sites did not use JavaScript API to map the data. Instead these sites display maps as images embedded in the HTML of the web pages. Figure~\ref{fig:gnb_ca} shows one example of this type of map. These maps do not use  XHR files to load the data. Instead, they load small markers in the HTML and overlay them on the map image. For these sites, clicking on the marker images will bring the user to another web page where the camera data is loaded.

\section{Automated Network Camera Discovery System}
\label{sec:methodology}
This section outlines our system which automatically discovers Network Camera data using common characteristics found from the sample websites in Section~\ref{sec:website_analysis}. Unlike methods described in our previous work~\cite{Dailey_imawmw}, the network camera data collection is done automatically. 

Previous work required a detailed analysis of the website structure to identify where the network camera data links appeared. For each website, the format and site organization described in Section~\ref{sec:website_analysis} would need to be manually determined. This work automates the discovery process using the similarities in the structure of network camera websites. The format and site organization categorizations in Section~\ref{sec:website_analysis} are used to identify patterns within the website that might contain network camera data. 

Instead of extracting only known network camera data links, we extract all data links that match the structures and formats outlined in Section~\ref{sec:website_analysis}. Then we filter out data links that match our definition of a network camera. By filtering out the data the method no longer depends on knowing where in the webpage structure the network camera data links are present.

The system has two parts: Section~\ref{subsec:meth_web_crawler} The Web Crawling module responsible for searching the website for potential Network Camera data links and Section~\ref{subsec:camera_id} The Network Camera Identification module that downloads and compares data from the data links found by the crawler.

\subsection{Web Crawler}
\label{subsec:meth_web_crawler}
\begin{figure}
	\centering
    	\includegraphics[width=.973\linewidth]{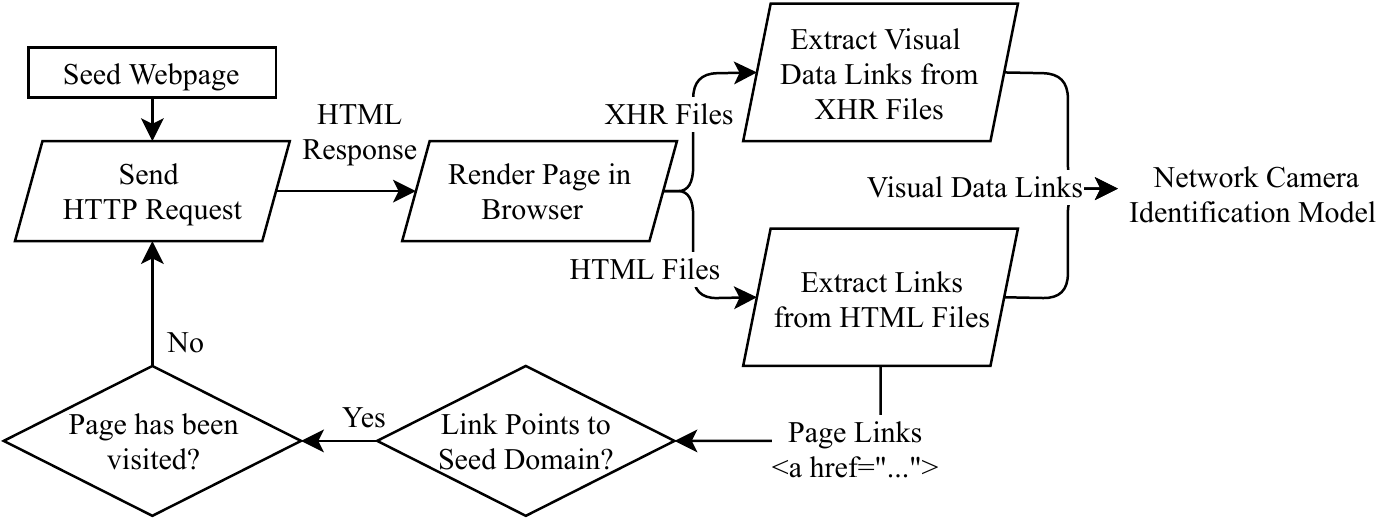}
    	\caption{The Web Crawler module: This module searches every page for links to other pages and visual data links. The Web Crawler module then sends visual data links to the Network Camera Identification module (Figure \ref{fig:camera_id_flowchart}).}
    	\label{fig:discovery_flowchart}
\end{figure}

If an IP address responds to an HTTP requests with a web page, then the address points to a website that may contain many Network Cameras. The flowchart in Figure~\ref{fig:discovery_flowchart} describes how the Web Crawler aggregates links to potential Network Camera data.

\begin{figure}
	\centering
	\begin{subfigure}{0.33\linewidth}
		\includegraphics[width=\textwidth]{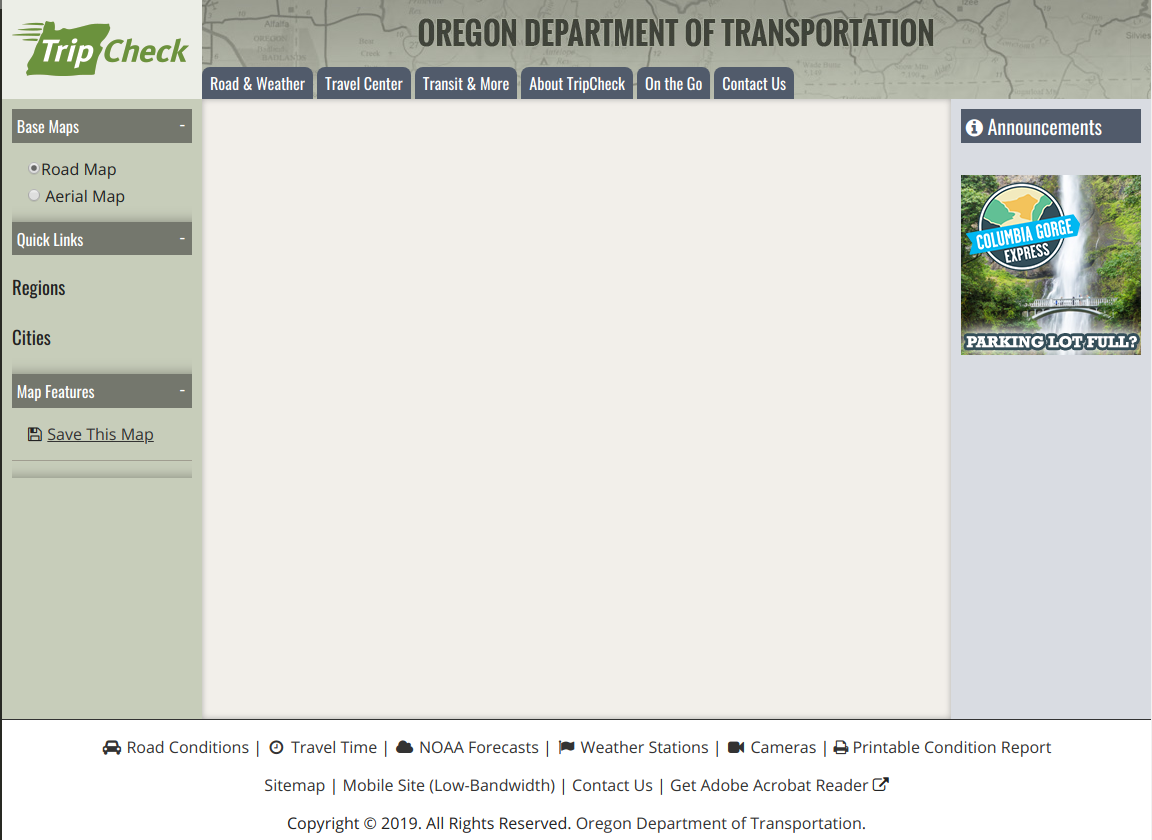}
		\caption{JavaScript disabled \label{fig:tripcheck_nojs}}
	\end{subfigure}
	\makebox[0.165\linewidth][c]{}
	\begin{subfigure}{0.33\linewidth}
		\includegraphics[width=\textwidth]{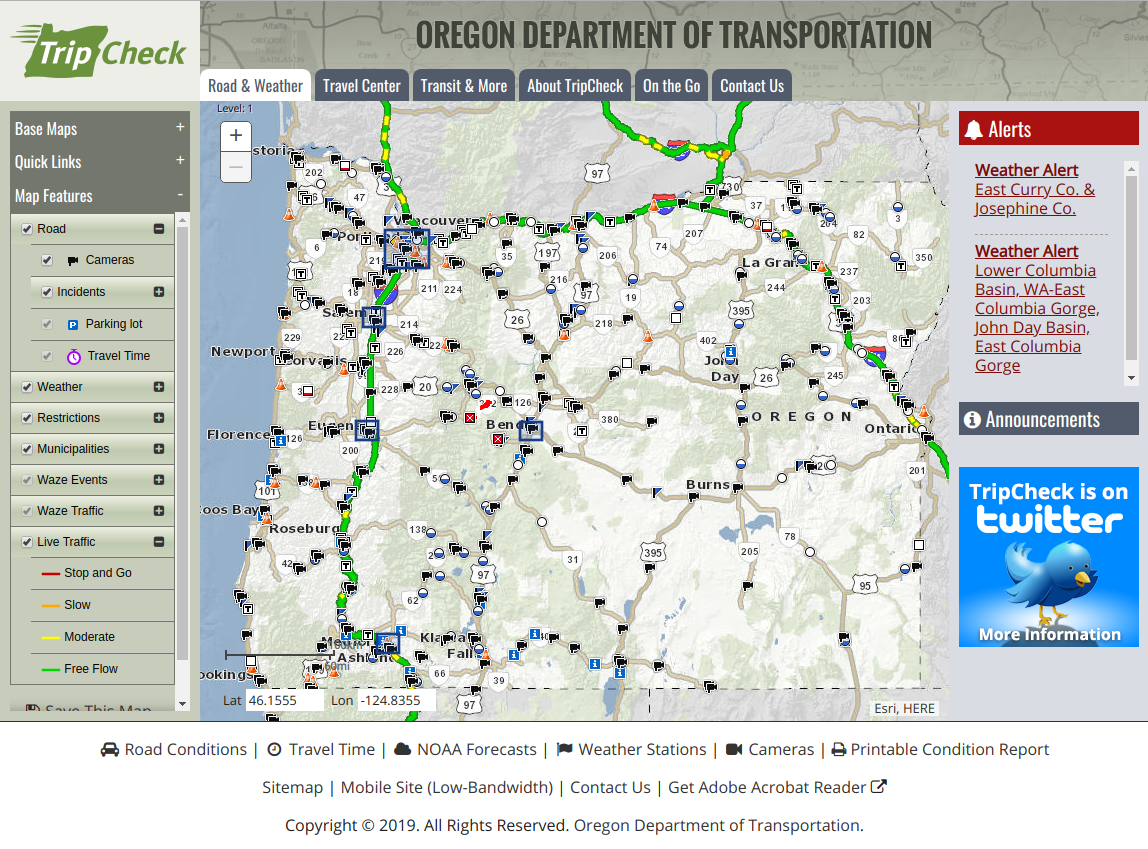}
		\caption{JavaScript enabled \label{fig:tripcheck_js}}
	\end{subfigure}
	\caption{Screenshots of Oregon Department of Transportation website~\cite{tripcheck} show the importance of JavaScript rendering for many Network Camera websites. If JavaScript rendering is not enabled, the Network Camera data in the map is not loaded.\label{fig:js_loading}}
\end{figure}

For each data link found on a web page, the Web Crawler module downloads the web page and parses the HTML. Some pages contain JavaScript code. Thus, each page is rendered in a browser environment to ensure all the page assets are loaded properly. Figure~\ref{fig:js_loading} shows that the rendering step is important because not executing the JavaScript can substantially reduce the information loaded into the web page. In many cases, Network Camera data is loaded into the site during this time using XHR requests as discussed in Section~\ref{sec:website_analysis}. Without this step, Network Camera data can potentially be lost. 

After the web page has been downloaded and rendered, the crawler will parse the HTML response and look for types of data links common to Network Cameras such as image and video links. The crawler will look for \verb|.jpg| or \verb|.png| file extensions, as these are the most common data formats found in Section~\ref{subsec:camera_formats}. For streaming cameras, the crawler will look for \verb|.m3u|, \verb|rtmp://|, \verb|rtsp://|, and \verb|.mjpg|  links. 

While the browser environment is loading the page, our system will monitor all XHR requests sent by the site. This will identify network camera data links loaded in the map views described in Section~\ref{subsec:map_views}. The Web Crawler module will then parse the request URLs to look for common map API database files with extensions such as \verb|.json|, \verb|.geojson|, and \verb|.xml|. These files are parsed for common Network Camera data links and links to  HTML pages. 

The Web Crawler module records all links found to ensure duplicate links are not checked twice. Links to other web pages are filtered to ensure the page has not been previously crawled and then sent to the crawler. 
At this stage, the extracted links have either been identified as new web pages and sent to the crawler, point to camera data formats (images or video), or discarded. The next step is to determine which of the data links are Network Cameras using the Network Camera Identification module.

\subsection{Network Camera Identification}
\label{subsec:camera_id}

\begin{figure}
	\centering
    \includegraphics[width=0.70\linewidth]{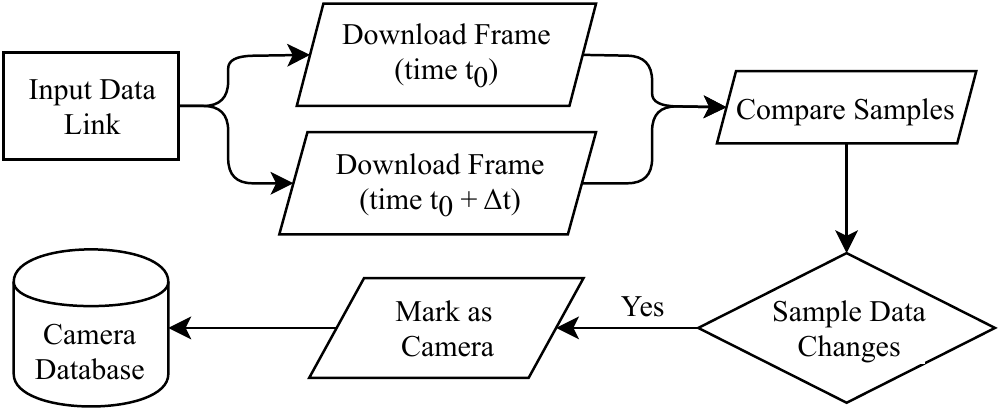}
    \caption{Network Camera Identification module: Each data link found by the Web Crawler module (Figure \ref{fig:discovery_flowchart}) is processed by this module to determine if a given link is to a Network Camera.}
    \label{fig:camera_id_flowchart}
\end{figure}

After potential Network Camera data links have been aggregated, the Network Camera Identification module determines if a given data link references Network Camera data by distinguishing between Network Cameras (Figure~\ref{fig:nyc_traffic}) and web assets (Figure~\ref{fig:web_assets}). The real-time data from Network Camera changes often; in contrast, web assets rarely change. For each data link, the Camera Identification module downloads several frames from the data link at different times. After each frame is downloaded, they are compared to determine if there is any change.

We present three different comparison methods for determining if a data link is a Network Camera. (1) \textit{Checksum Method}: compare the file checksum across sample frames downloaded by the Network Camera Identification Module. (2) \textit{Percent Difference Method}: compare the percent of pixels changed between frames. (3) \textit{Luminance Difference Method}: compare the mean pixel luminance change between frames. Each of these comparison methods are evaluated on a manually labeled dataset in Section~\ref{subsec:id_metrics}.

The first and simplest method is to compare the checksum on the frames. For this method, the Network Camera Identification Module simply compares the MD5 checksum between different frames taken from the visual data link. If the checksum changes between the frames taken at $t_0$ and $t_0 + \Delta t$ we classify the data link as a Network Camera. Although this method is simple to implement, it does not look at the content of the frame to predict if it came from a Network Camera. The next two methods we use statistics that provide more information about the content of the frame. 

The Percent Difference Method compares the percentage of pixels that change between the frames. Algorithm~\ref{alg:percentDiff} shows how this value is calculated. To determine if a visual data link is from a Network Camera, a threshold value is determined experimentally. If the percentage of pixels that changed is above the threshold, we classify the visual data link as a Network Camera. This method provides more insight to how the content of the visual data link changes between sample frames. For each frame, a small percentage change may indicate that only a few pixels changed. Real-time data changes significantly over time. This would not be true of computer generated images like those in Figure~\ref{fig:web_assets}. Unlike the Checksum Method, this method takes the content of the frame into account.

The final method we introduce is the Luminance Difference Method. This method uses the average difference in luminance between sample frames. For this method, shown in Algorithm~\ref{alg:luminance}, the mean pixel value of the frame is taken. The difference between the luminance of the initial frame ($t_0$) and a second frame ($t_0 + \Delta t$) is used to identify Network Cameras. We determine an experimental threshold to classify Network Cameras in Section~\ref{subsec:id_metrics}. If the overall luminance difference between the sample frames is greater than this threshold, we classify the visual data link as a Network Camera.

This method uses the fact that many Network Cameras are positioned outdoors. Network Cameras will have a greater difference in luminance over the course of a day as the sun changes position. This should also provide some insight into the content of the image and improve the overall accuracy of our classification. 

Streaming camera links aggregated by the Web Crawler module are processed differently by the Network Camera Identification module. For these data links, the module will attempt to establish a connection to the stream. The method used to establish this connection changes depending on the type of stream. For example, if the link is to an HLS stream, the Camera Identification module will send an HTTP request to download the \verb|.m3u| playlist file. Using this playlist file, the module will connect to the stream and download camera data. The Camera Identification module will determine if the link is to a real-time data stream by checking the \verb|duration| and \verb|start time| information for the video. The way this information is obtained depends on the streaming format used. A streaming camera will have a \verb|start time| greater than zero and no \verb|duration|. If the data link passes these checks, it is considered a Network Camera. Frames can also be taken from streaming cameras that pass the first checks. These frames can be compared using one of the three methods for non-streaming cameras. 

\begin{algorithm}
	\caption{Percentage Difference}\label{alg:percentDiff}
	\begin{algorithmic}
	   \Procedure{percentDiff}{img1, img2}
	        \State count $\gets 0$
	        \For{px $= 0,1,\dots, $\Call{size}{img1}}
	            \State count $\gets$ count$+ |$img1[px] - img2[px]$| > 0$
	        \EndFor
	        \State \Return count$/$\Call{size}{img1}   
	   \EndProcedure
	   \If{\Call{percentDiff}{frame0,frame1} $>$ threshold}
            \State \Return True \Comment{visual data link is a Network Camera}
      \Else
            \State \Return False \Comment{visual data link is not a Network Camera}
      \EndIf
    \end{algorithmic}
\end{algorithm}

\begin{algorithm}
	\caption{Luminance Difference}\label{alg:luminance}
	\begin{algorithmic}
	    \Procedure{luminanceDiff}{img1, img2}
    	    \State img1Lum $\gets$ \Call{mean}{img1}
    	    \State img2Lum $\gets$ \Call{mean}{img2}
        \State \Return \Call{abs}{img1Lum - img2Lum}
        \EndProcedure
        \If{\Call{luminanceDiff}{frame0, frame1} $>$ threshold}
            \State \Return True \Comment{visual data link is a Network Camera}
        \Else
            \State \Return False \Comment{visual data link is not a Network Camera}
        \EndIf
	\end{algorithmic}
\end{algorithm}

\section{Implementation}
\label{sec:implementation}
This section outlines the implementation of the proposed system for automated Network Camera discovery.

\subsection{Web Crawler Implementation}

The Web Crawler Module is implemented using the Scrapy web crawler framework ~\cite{scrapy} and uses Splash web browser \cite{splash} to render the JavaScript on the crawled web pages. The aggregated meta-data found by the crawler is stored in a MongoDB~\cite{mongodb} unstructured database.

The Splash rendering engine is responsible for sending the HTTP requests to the target web page. If Splash could not establish a connection to a page within 180 seconds, the page is discarded and the crawler moves on. After the crawler connects to the server and downloads the web page, the JavaScript render engine waits for 8 seconds for the page to finish loading. This step ensures JavaScript assets are loaded into the page before the rendered HTML is sent to the HTML parser.

The HTML parser finds new pages within the seed domain by extracting the \verb|<a href="...">| tags. The contents of these tags link to other web pages. The crawler is restricted to following \verb|<a href="...">| tags that link to same domain. For example, if the seed website has the domain of \verb|www.example.com| then the crawler would follow links to \verb|www.example.com/cameras/| and \verb|www.subdomain.example.com/cameras/| but would not follow links to \verb|www.facebook.com/|. This restriction prevents the crawler from spending time crawling large websites outside the target domain. In addition, as several sites discussed in Section~\ref{sec:website_analysis} do not host or embed camera data on their servers, this restriction on the crawler will also prevent it from finding camera data that is not embedded on the target domain. This restriction could be lifted for future crawls enabling the crawler to discover new domains that host Network Camera data. 

In this implementation, the parameters given in each web server's \verb|robots.txt| file are respected. If the \verb|robots.txt| file does not exist for a domain or if the site did not have specific crawling directives, the following default parameters are used:

\begin{itemize}
    \item The crawler is limited to 32 connections per domain so as not to overwhelm the web servers. 
    \item The crawler waits 3 seconds between requests to the same target domain. 
\end{itemize}

The crawler traverses sites in breadth first order starting on the seed page and reaching a maximum depth of 15 pages from the seed page. Once the crawler has visited all the available links within the seed domain or has reached a depth of 15, the crawler stops crawling that domain. The crawler keeps detailed information on the data aggregated. In addition to the absolute links to the images discovered by the crawler, the HTML data of each web page and other statistics are stored in the Crawler Database. 

\subsection{Camera Identification Module Implementation}
The Camera Identification Module takes the links discovered by the Web Crawler Module and determines whether or not they contain Network Camera data. When a new data link is aggregated by the crawler, the Camera Identification Module downloads a copy of that image and stores it in the database. The Identification Module downloads a frame from each data link 4 times.

Several different algorithms were tested to determine if an image was from a Network Camera. Detailed discussion of the luminance and other methods tested can be found in Section~\ref{subsec:id_metrics}. 
In total, 4 frames were downloaded from each data link at the following times:
\begin{itemize}
    \item[] $t_0$ (Time when the data link is first aggregated by the Web Crawler)
    \item[] $t_0 + 5min$
    \item[] $t_0 + 60min$
    \item[] $t_0 + 12hrs$
\end{itemize}
The range of time from the first frame at time $t_0$ to time $t_0 + 12hrs$ is chosen to provide a significant change in luminance that is often seen in outdoor Network Cameras.
In the next section we discuss the results of this implementation on the sample of 73 example websites introduced in Section~\ref{sec:website_analysis}.

\section{Results}
\label{sec:results}

\begin{figure}
	\centering
	\includegraphics[width=0.75\linewidth]{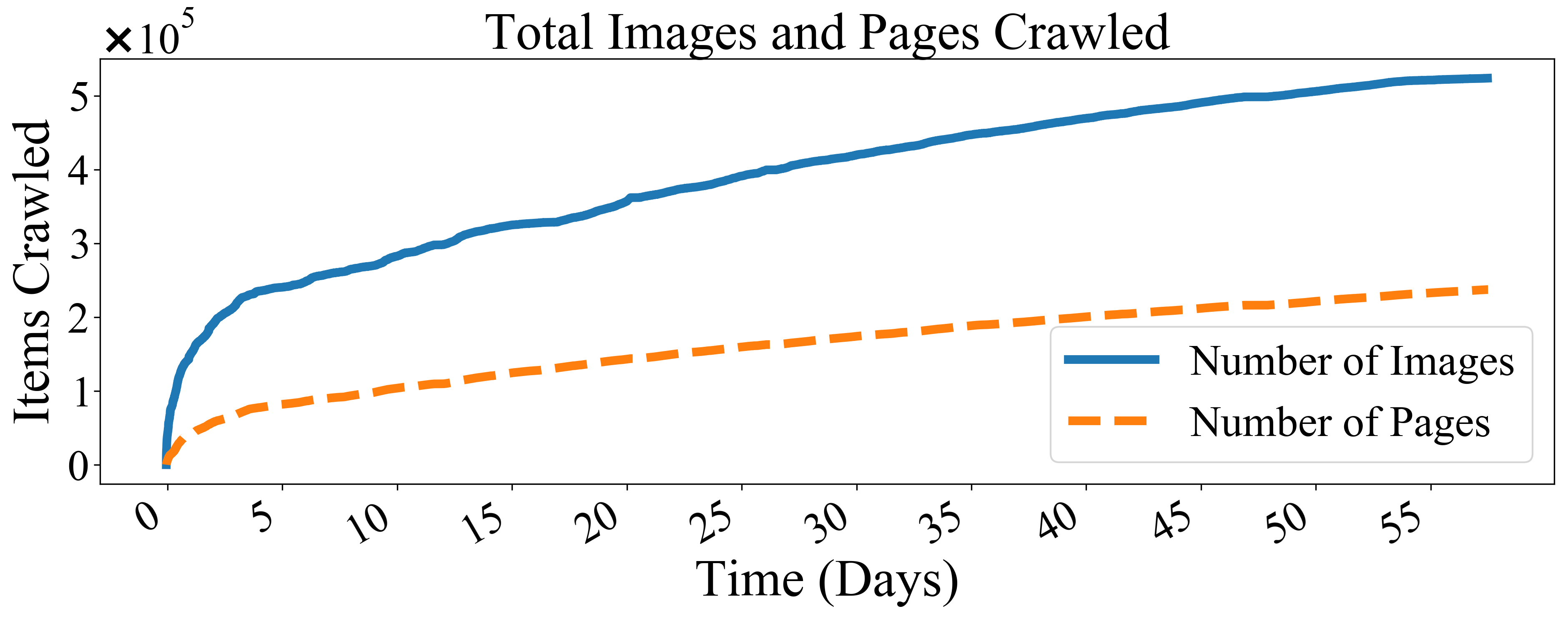}
	\caption{Total number of images and pages found during the Network Camera Web Crawler experiment. 
	\label{fig:items_by_time}}
\end{figure}

The Network Camera Discovery module is tested on the sample of 73 Network Camera sites introduced in Section~\ref{sec:website_analysis}. We run the Network Camera Discovery module for 55 days and find 523,696 visual data links. During the run, the module crawls 237,257 unique web pages in total. Figure~\ref{fig:items_by_time} shows the total number of pages and visual data links found by the crawler during the run. 
Of the 73 sample websites, 4 websites were not crawled successfully. This error is because the web pages are unreachable at the start of the crawl, and no new links are generated.

\begin{figure}
    \centering
    \includegraphics[width=0.75\linewidth]{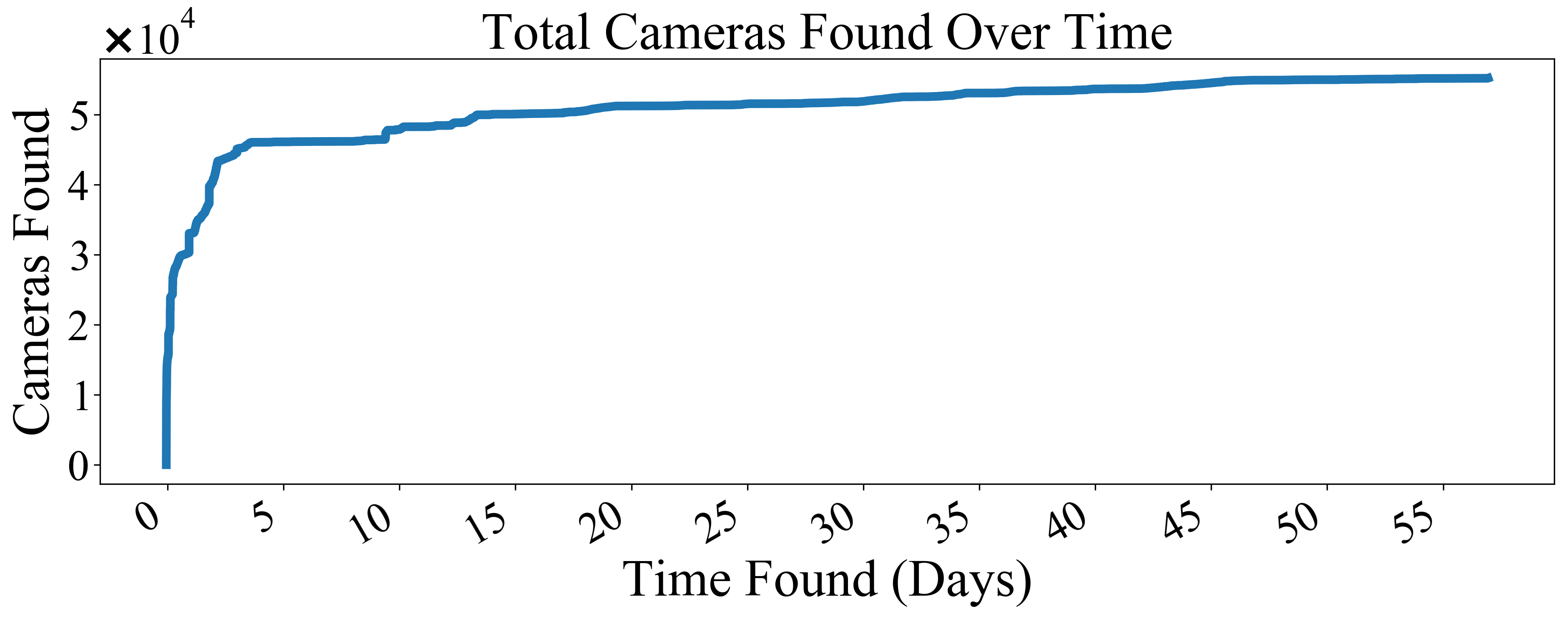}
    \caption{The number of cameras found over the length of the crawl.}
    \label{fig:when_cams_found}
\end{figure}

Potential cameras are found on 47 of the 73 websites. Figure~\ref{fig:when_cams_found} shows when different cameras are found during the crawl. Most cameras are found at the beginning of the crawl, because the Web Crawler starts on the page linking directly to the camera data for most sites.

A breakdown of the data links found for streaming cameras can be found in Table~\ref{tab:streaming_images_found}. Of the data links found, less than 1\% are links to video data. Only 3,974 video data links are discovered by the crawler and 99\% of these links are found in XHR files. Only 25 cameras are found to be embedded in the page HTML. All 25 of the embedded links are for \verb|.mjpg| cameras and 17 of the 25 are working Network Cameras. Of the 3,949 streaming data links found in XHR files, about 43\% are links to Network Cameras. In total 1,745 streaming Network Cameras are found from 8 different websites. These streaming Network Cameras were validated using a combination of the \verb|start time| and  \verb|duration| check and the Luminance Difference method described below with the optimal experimental threshold.

Table~\ref{tab:static_images_found} shows where the Network Camera data links were found. Of the 465,249 unique image links found, 89\% are embedded in the HTML files, and the remaining 11\% are found in the XHR files. The images found in XHR files are much more likely to be Network Cameras then those embedded in the HTML pages. Almost 30\% of images found in XHR files are found to be from Network Camera while less than 4\% of HTML embedded images come from Network Cameras. Static image camera data is found using the average image luminance method, which is discussed in further detail in the Section~\ref{sec:analysis}.

In some websites, more cameras are found by the Camera Discovery module than manual estimation. In some cases i.e. the California Department of Transportation site \cite{dot_ca_gov}, the site contains links to static camera images and streaming links, both of which would be caught by the Camera Discovery module. In other cases, for example the USGS geological monitoring site \cite{hvo_wr_usgs}, each camera image has an associated thumbnail image posted to a different URL. More examples and analysis of the sites can be found in Section~\ref{sec:analysis}.

\begin{figure}
    \centering
    \includegraphics[width=0.5\linewidth]{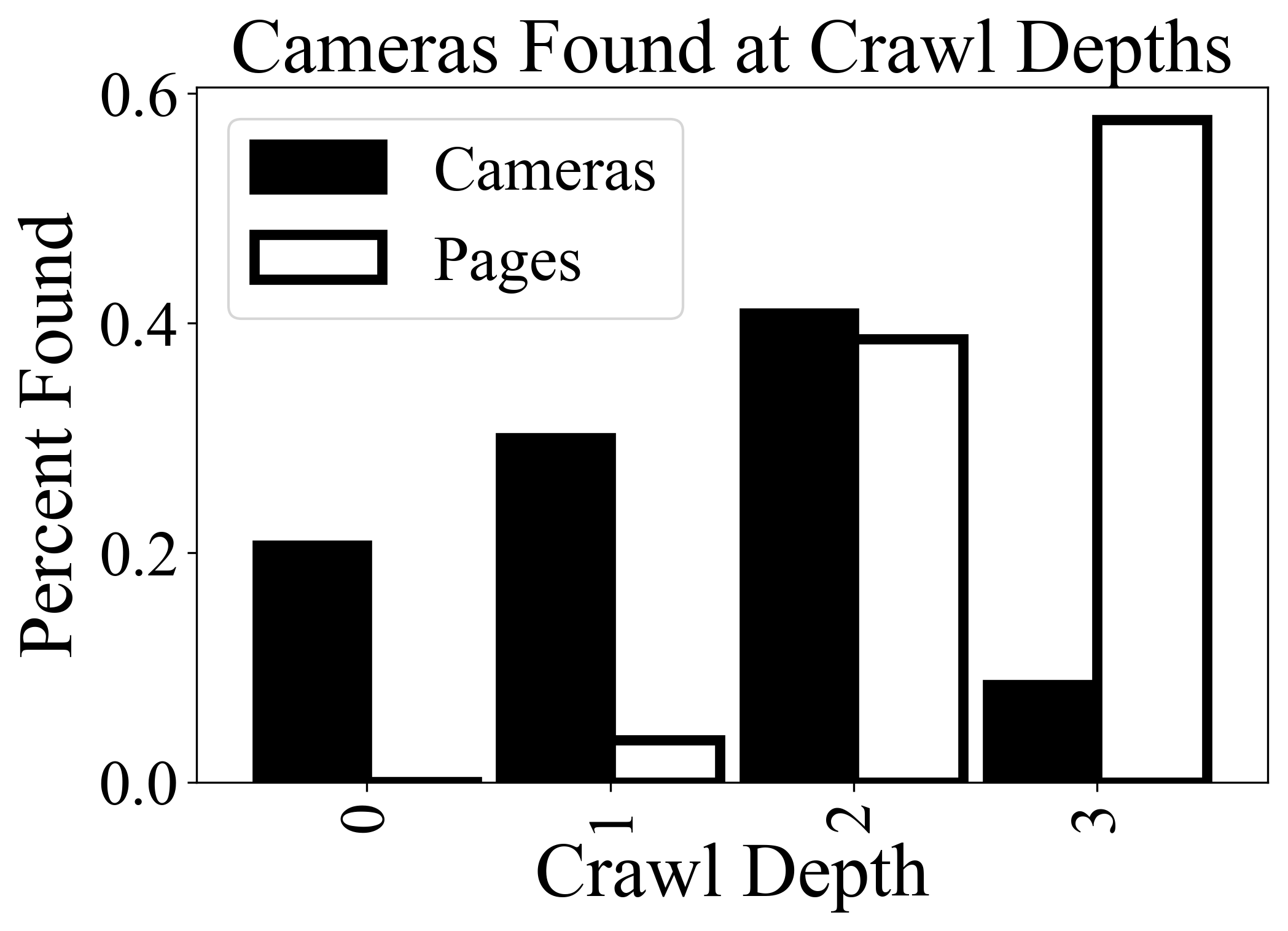}
    \caption{The percentage of cameras and pages found at each page depth during the crawl. }
    \label{fig:crawl_depth}
\end{figure}

\begin{table}
    \centering
    \begin{tabular}{l|r|r}
        \hline
        \textbf{Static Images} & \textbf{Num. of Links found} & \textbf{Num. of Cameras found} \\ \hline
        From HTML & 466,983 & 16,998 \\
        From XHR & 55,636 & 15,960 \\ \hline
        Total & 522,619 & 32,958 
    \end{tabular}
    \caption{Static images and cameras found during the crawl.}
    \label{tab:static_images_found}
\end{table}

\begin{table}
    \centering
    \begin{tabular}{l|r|r}
        \hline
        \textbf{Streaming} & \textbf{Num.~of Links found} & \textbf{Num.~of Cameras found} \\ \hline
        From HTML & 25 & 17 \\
        From XHR & 3,949 & 1,728 \\ \hline
        Total & 3,974 & 1,745  \\
    \end{tabular}
    \caption{Streaming videos and cameras found during the crawl.}
    \label{tab:streaming_images_found}
\end{table}

Figure~\ref{fig:crawl_depth} shows the crawler does not traverse websites very deeply and only reached a max depth of 3. In this implementation, the scheduler preformed a breadth first traversal of each site. On our target sites, pages containing network cameras could be reached within a traversal depth of 3 in most cases. Although many more pages are crawled at depth 3, a majority of the cameras are found at a depth of 2. 

\section{Analysis}
\label{sec:analysis}

In this section, we will analyze the results of our system in more detail. In Section~\ref{subsec:id_metrics}, we discuss methods tested to differentiate between Static Image Cameras and other web assets. In Section~\ref{subsec:website_examples}, we give a few in-depth examples of the successes and failures of the Camera Discovery module.

\subsection{Network Camera Identification Metric Analysis}
\label{subsec:id_metrics}
This section discusses different methods that can be used to differentiate between Network Cameras and other images aggregated by the Web Crawler. The three methods presented in Section~\ref{subsec:camera_id} are tested using precision and recall. Precision is a measure of proportion of Network Cameras correctly identified by our system (true-positives) divided by the total number of cameras identified (true-positives + false-positives). Recall is the percentage of cameras correctly identified by our system (true-positives) divided by the total number of cameras (true-positives + false-negatives). 

We show that the luminance change method has the highest precision and recall on the test dataset. The test dataset is created using a subset randomly chosen from data aggregated by the Web Crawler. For each data link, 4 frames are taken by the Network Camera Identification module. These frames are then manually identified by a person. 

\begin{figure}
    \centering
    \newcommand{\imgwidth}{0.32\linewidth}
    \newcommand{\simgwidth}{0.30\linewidth}
    
    \newcommand{\imheader}{
        \makebox[\imgwidth][c]{{$t_0$}}
        \makebox[\imgwidth][c]{{$t_0 + 1$hr}}
        \makebox[\imgwidth][c]{{$t_0 + 12$hrs}}
    }
    \begin{subfigure}{0.55\linewidth}
        \centering
        \includegraphics[width=\imgwidth]{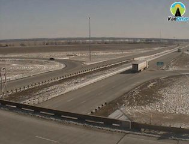}
        \includegraphics[width=\imgwidth]{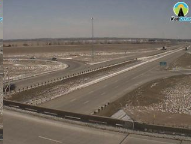}
        \includegraphics[width=\imgwidth]{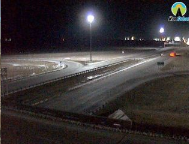}
        \caption{\label{fig:labeled_cameras}}
    \end{subfigure}
    \begin{subfigure}{0.20\linewidth}
        \centering
        \includegraphics[width=\simgwidth]{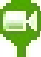}
        \includegraphics[width=\simgwidth]{figures/label_examples/Bad_1-123.png}
        \includegraphics[width=\simgwidth]{figures/label_examples/Bad_1-123.png}
        \caption{\label{fig:labeled_noncameras1}}
    \end{subfigure}
    \begin{subfigure}{0.20\linewidth}
        \centering
        \includegraphics[width=\simgwidth]{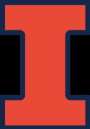}
        \includegraphics[width=\simgwidth]{figures/label_examples/Bad_2-123.png}
        \includegraphics[width=\simgwidth]{figures/label_examples/Bad_2-123.png}
        \caption{\label{fig:labeled_noncameras2}}
    \end{subfigure}
    \caption{Examples of labeled data. Each set of images (from left to right) shows: initial frame taken, 5min after, 60min after, and 12hours after. Images in (a) show links that were labeled \textit{Network Camera}. 
    (b) and (c) links labeled \textit{Other Web Assets}.}
    \label{fig:label_examples}
\end{figure}

\begin{figure}
    \centering
    \begin{subfigure}{0.49\linewidth}
        \includegraphics[width=0.49\linewidth]{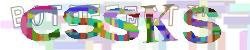}
        \includegraphics[width=0.49\linewidth]{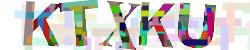}
        \caption{}
        \label{fig:CAPTCHA_error}
    \end{subfigure}
    \begin{subfigure}{0.49\linewidth}
        \includegraphics[width=0.49\linewidth]{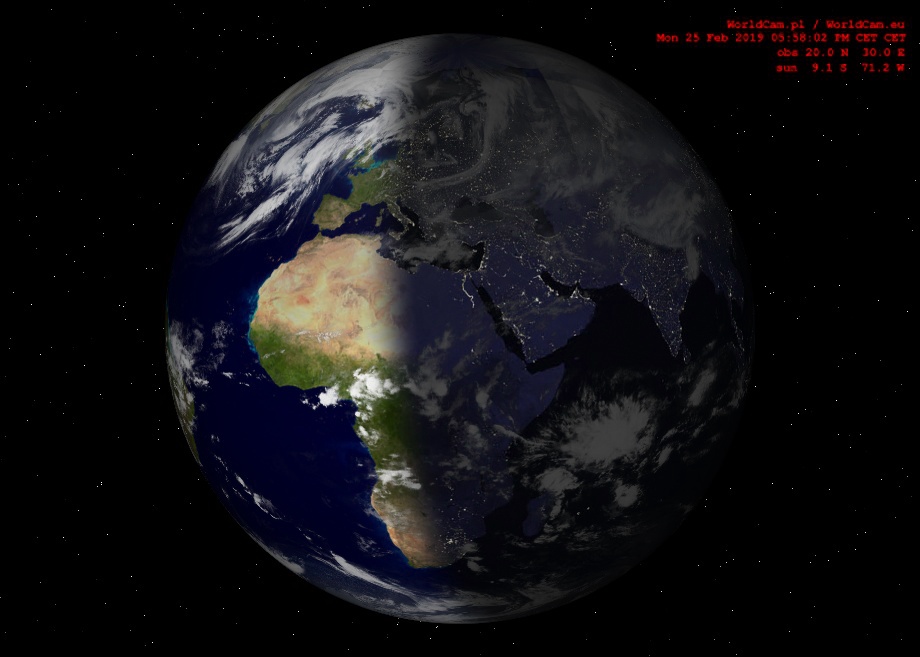}
        \includegraphics[width=0.49\linewidth]{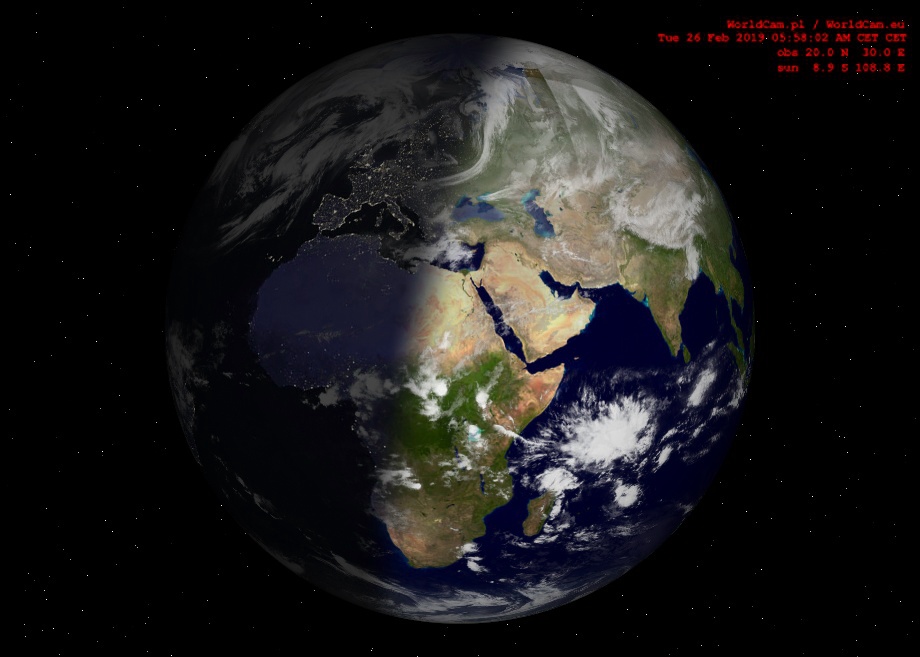}
        \caption{}
        \label{fig:CGI_error}
    \end{subfigure}
    \caption{Above are examples of images incorrectly classified as Network Cameras using the Luminance Difference Method. More sophisticated methods are needed to identify these false positive classifications ~\cite{worldcam_eu}.}
    \label{fig:error_examples}
\end{figure}

Each set of frames from one data link is given a single label of one of the following categories:
\begin{enumerate}
    \item \textit{Network Camera} - Any chosen data link where images change between frames. The frames from the link also have to resemble Network Camera images. See Figure~\ref{fig:label_examples}a.
    \item \textit{Other Web Assets} - Any other data link where sample frames do not resemble a Network Camera. See Figure~\ref{fig:label_examples}c. In some cases the web assets change over time but the data does not appear to be from a Network Camera. Figure~\ref{fig:error_examples} shows examples of such data.
\end{enumerate}

\begin{figure}[h]
    \centering
    \begin{subfigure}{0.45\linewidth}
        \centering
        \includegraphics[width=\textwidth]{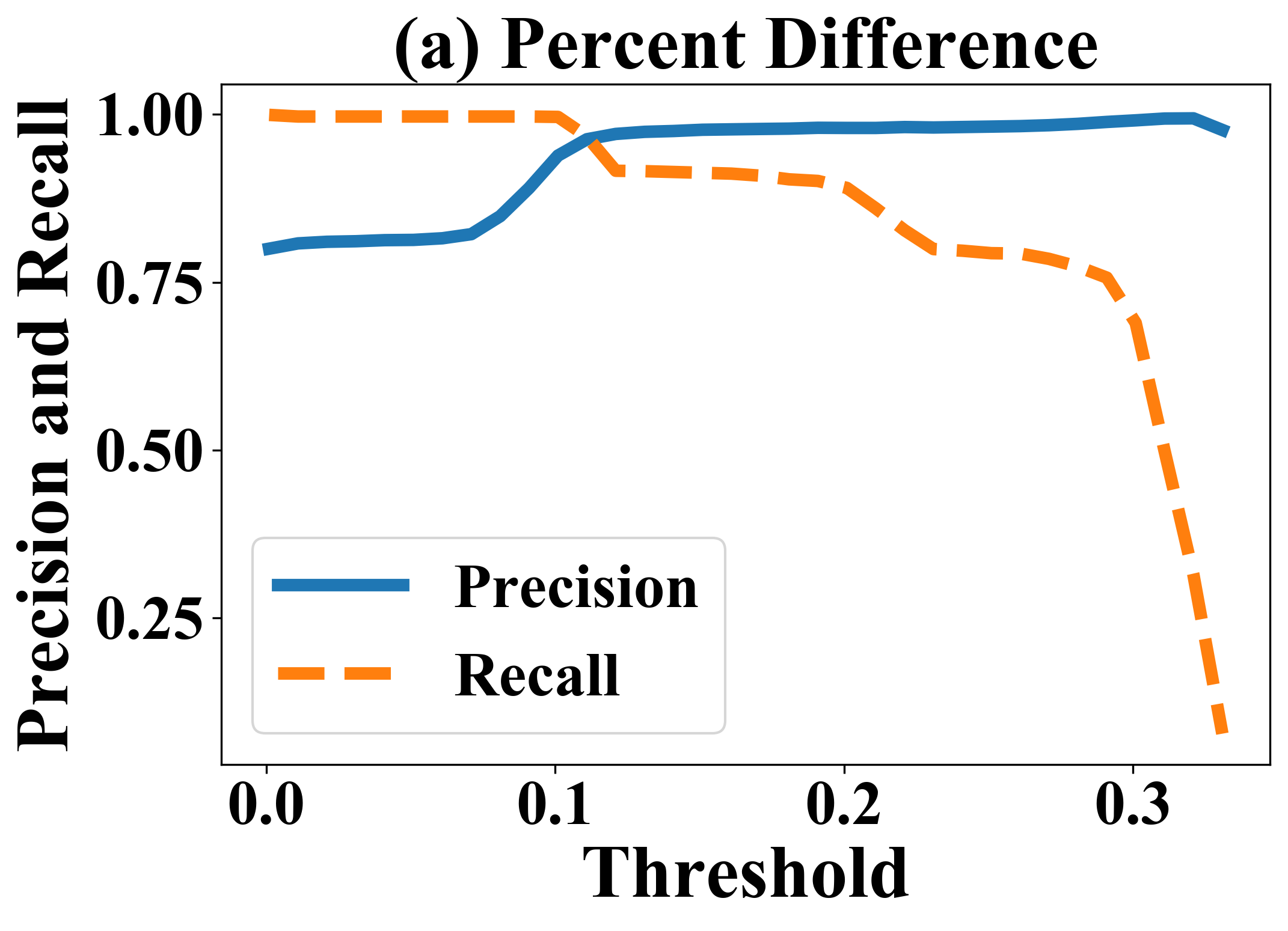}
    \end{subfigure}
    \makebox[0.033\linewidth][c]{}
    \begin{subfigure}{0.45\linewidth}
        \centering
        \includegraphics[width=\textwidth]{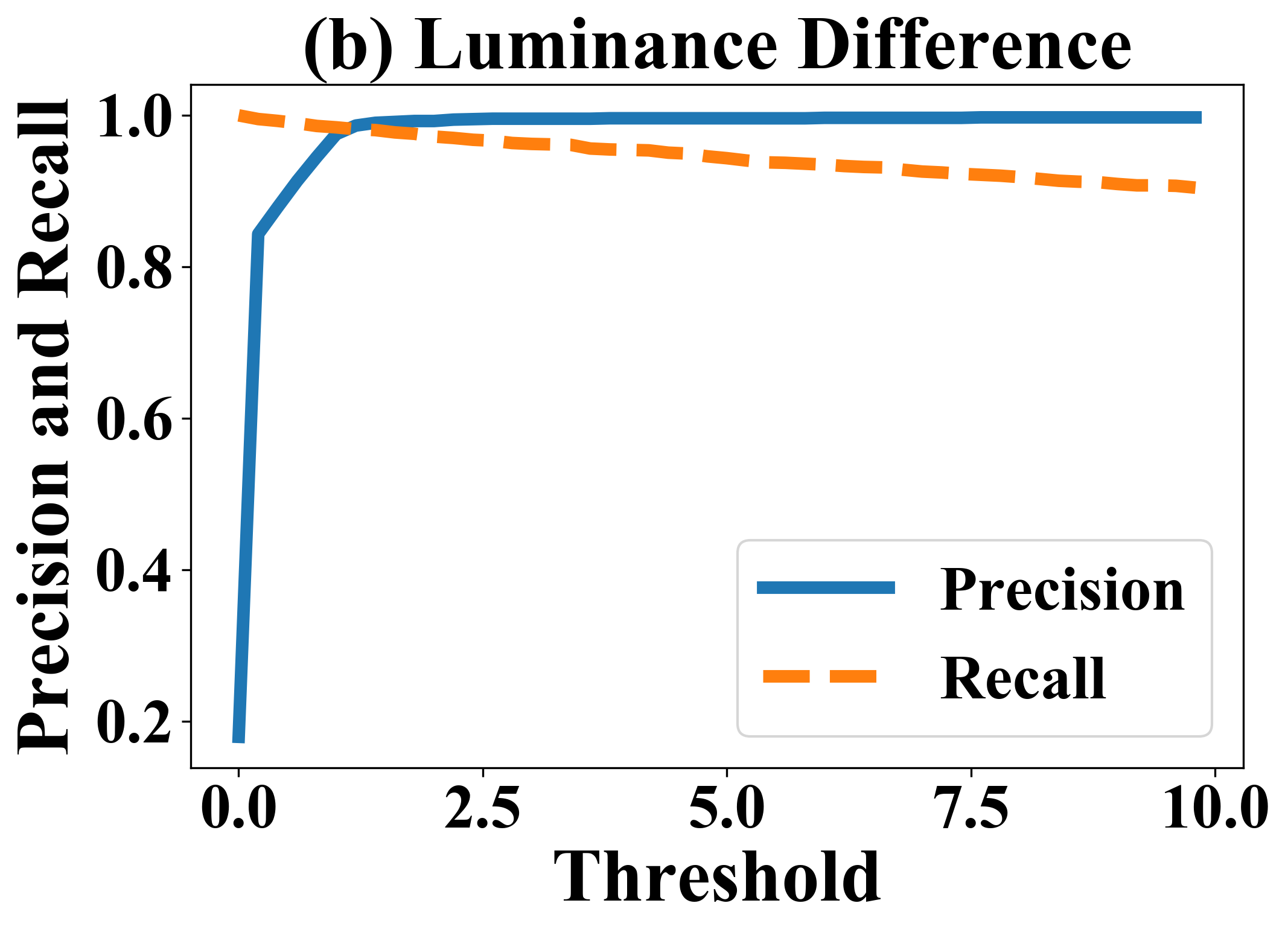}
    \end{subfigure}
    \caption{Figure \ref{fig:PrecisionRecall_PcntDiff_and_Luminance}a shows the precision and recall evaluated for the percentage difference method presented in Section~\ref{subsec:camera_id}. The threshold of this method is the average percentage of pixels that changed over the sample frames. Figure \ref{fig:PrecisionRecall_PcntDiff_and_Luminance}b shows the precision and recall of the Luminance Difference Method from Section~\ref{subsec:camera_id}. Both methods are evaluated on the set of manually labeled data. \label{fig:PrecisionRecall_PcntDiff_and_Luminance}}
\end{figure}

During the labeling process, the person is shown all the frames taken by the Network Camera Identification module side by side, along with an indication of the number of pixels that changed between the frames. If no pixels changed over the from frame $t_0$  to frame $t_0 + 12hours$, the data link could not be labeled as a Network Camera. The person labeling the image would then look for large changes in light level over the sample period. If the image appeared to be taken from a static camera, was not computer generated, and had visible changes to the scene such as light level change, the image would be labeled as a Network Camera. If the changes were too subtle to see across the samples, the data link could also be analyzed for words such as "webcam", "camera", etc.

In total, 9,165 images are manually labeled. This sample represents about 2\% of the total visual data links aggregated by the crawler. In the labeled dataset 1,645 (18\%) of the images are found to be from Network Cameras. 

The remaining 7,520 (82\%) of images are other web assets.

The first method outlined in Section~\ref{subsec:camera_id}, the Checksum Method, uses only the checksum of the frame to determine if the data has changed. This method identifies 818 of the labeled images as Network Cameras. The recall of this method is 100\% because an image has to change to be classified as a Network Camera. However, there is a high number of false positives (images that changed but were not from Network Cameras); as a result, the precision is 75.0\%.

The Percent Difference Method outlined in Section~\ref{subsec:camera_id} finds the percentage of pixels that change between the frames. For this method, we use a linear combination of the differences from each frame relative to the initial frame. This average difference across frames is used to determine a threshold. Figure~\ref{fig:PrecisionRecall_PcntDiff_and_Luminance}a shows a graph of the precision recall over different thresholds. 

If we treat precision and recall as having equal importance for our classifier, the threshold is 0.11. This threshold identifies 1,683 of the labeled data links as Network Cameras. The overall accuracy of this method for the manually labeled dataset was 99.1\%. The precision and recall of this method is 96.3\% and 98.5\% respectively. If we apply this method to all the visual data links found by the Web Crawler we classify 55,365 visual data links as Network Cameras. 

The Luminance Difference Method from Section~\ref{subsec:camera_id} was found to be the most accurate of the three methods tested on the labeled dataset. For this method, the mean pixel value of the image is taken from the $t_0$ frame and subtracted from the frame at $t_0 + 12~hours$. The resulting value is compared to an experimental threshold. The frame taken at 12 hours is chosen to capture the day/night cycle of outdoor cameras. 

If we treat precision and recall as having equal value, the threshold for the Luminance Difference Method is 1.3. At this threshold, this method has a precision of 98.7\% and a recall of 98.2\% on the labeled dataset. A graph of the precision and recall for this method can be found in Figure~\ref{fig:PrecisionRecall_PcntDiff_and_Luminance}b. If we apply this method with a threshold of 1.3 to the entire set of visual data links found by the Web Crawler Module, the total number of cameras identified is 55,619. 

\begin{figure}[t]
    \centering
    \includegraphics[width=\linewidth]{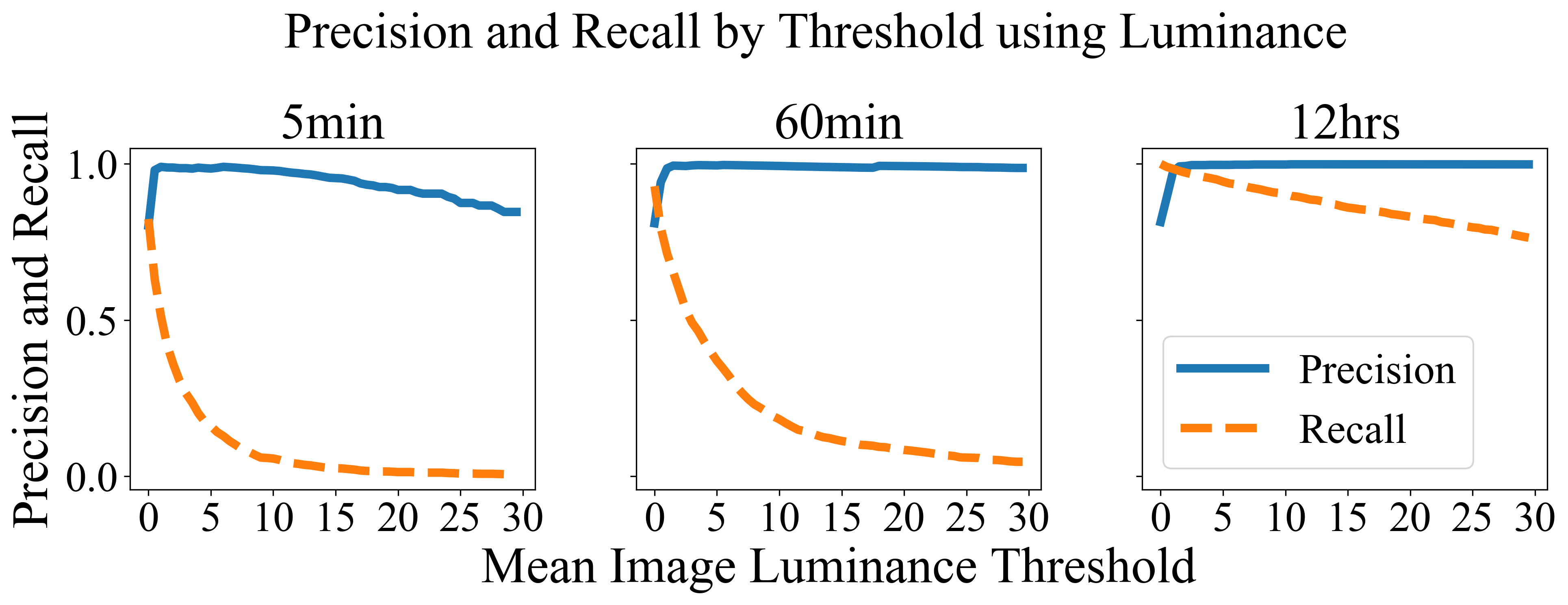}
    \caption{Comparison of the precision and recall using the mean Luminance Difference between the first frame and frames downloaded 5min after, 60min after, and 12hrs after the first frame.}
    \label{fig:prec_rec_byImage}
\end{figure}

A full comparison of the results of the three methods can be found in Table~\ref{tab:classifier_results}. Here we see that the Luminance Difference Method has the best accuracy on the labeled dataset.  Figure~\ref{fig:prec_rec_byImage} shows the precision and recall for the Luminance Difference Method for the frames taken at 5min, 60min, and 12hrs after the initial image was discovered. The difference between frame $t_0$ and frame $t_0 + 12~hours$ has the best trade off between precision and recall. This is likely due to the large change in luminance from day to night over a 12 hour period.

Table~\ref{tab:classifier_results} also shows the number of visual data links that are classified as Network Cameras for each method evaluated on the manually labeled dataset. We see that the Checksum Method labeled the most visual data links as Network Cameras. The manually labeled dataset contained 1,645 Network Cameras but the Checksum Method identified 2,061 visual data links as Network Cameras. This means that 416 visual data links that changed in the test dataset were not from Network Cameras. The Percent Difference Method and the Luminance Difference Method had a better overall precision meaning that the content of the frames can be used to help identify Network Cameras. 

We select the Luminance Difference Method as the best method for our classifier and apply this method to all visual data links collected by the Web Crawler Module. We find that 55,619 Network Cameras are identified. We add the 1,745 streaming Network Cameras to determine a final total of 57,364 Network Cameras. This number is close to the manually estimated number of cameras in Section~\ref{sec:website_analysis}. We will further discuss the results on specific sample websites in Section~\ref{subsec:website_examples}. Next, we look at some examples of visual data links that are incorrectly classified by the Luminance Difference Method.

\begin{table}[]
\centering
\small
\begin{tabular}{lrrrrr}
\hline
\multicolumn{1}{c}{} & \multicolumn{1}{c}{\textbf{Precision}} & \multicolumn{1}{c}{\textbf{Recall}} & \multicolumn{1}{c}{\textbf{Accuracy}} & \multicolumn{1}{c}{\textbf{\begin{tabular}[c]{@{}c@{}}Num.~of \\ Cams.~Found\\ (Labeled)\end{tabular}}} & \multicolumn{1}{c}{\textbf{\begin{tabular}[c]{@{}c@{}}Num.~of \\ Cams.~Found\\ (All Links)\end{tabular}}} \\ \hline
\textbf{Checksum} & 0.750 & 1.000 & 0.978 & 2,061 & 71,471 \\
\textbf{\% Diff.} & 0.963 & 0.985 & 0.991 & 1,683 & 55,365 \\
\textbf{Lum. Diff.} & 0.987 & 0.982 & 0.994 & 1.637 & 55,619 \\ \hline
\end{tabular}
\caption{Comparison of the results of the Network Camera classification methods introduced in Section~\ref{subsec:id_metrics}. For the Percent Difference Method, a threshold of 0.11 was used. For the Luminance Difference Method, a threshold of 1.3 was used. The number of visual data links classified as Network Camera using each method is shown for the manually labeled dataset. The last column shows the number of cameras that are found if each method is applied to all visual data links found by the Web Crawler Module.}
\label{tab:classifier_results}
\end{table}

\begin{figure}
	\centering
	\includegraphics[width=0.5\linewidth]{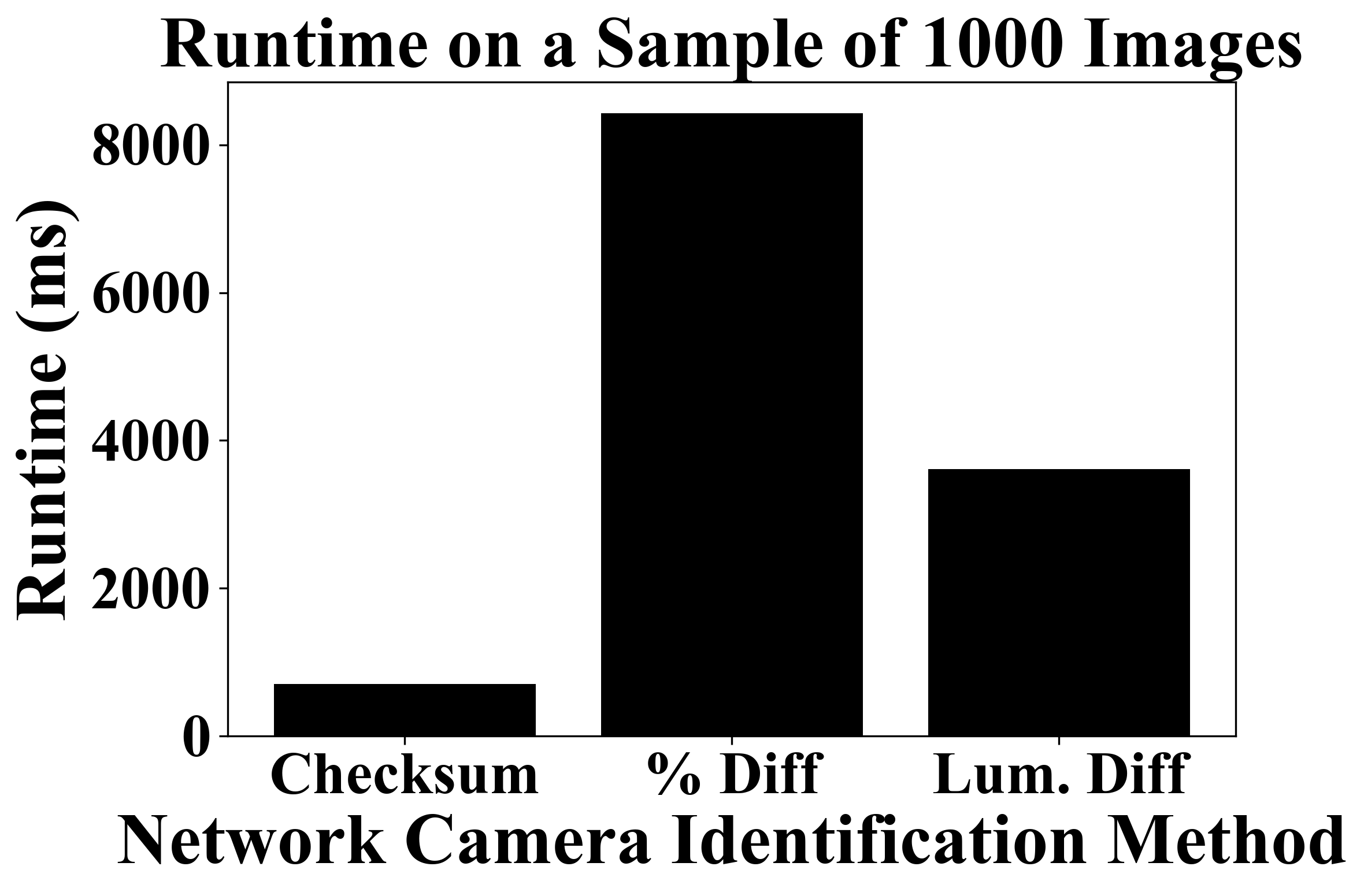}
	\caption{Runtime on a sample of 1000 images of the proposed network camera identification methods. 
	\label{fig:runtimes}}
\end{figure}

Figure~\ref{fig:error_examples} shows some examples of false-positive classifications of the image luminance method. CAPTCHA errors (Figure~\ref{fig:CAPTCHA_error}) are a common false-positive classification for the change in luminance. These images are hard to classify correctly without more advanced computer vision metrics. A few computer generated images are also incorrectly classified as Network Cameras by the luminance metric. Figure~\ref{fig:CGI_error} shows one such example that is updated to the same URL like a Network Camera, but instead depicts the movement of the sun on the earth. 
In addition to the accuracy of the methods, the computational cost of the 3 proposed methods was also analyzed. The results of this analysis can be seen in Figure~\ref{fig:runtimes}. Each method was timed on the same machine on a set of 1000 sample images. For each method, the average was used across 20 runs of the same data. The Checksum Method was the fastest of the proposed methods and was more then 5 times faster then the Luminance Difference Method. The average runtime of the chosen method for identifying network camera data links (the Lumianance Difference method) was found to be 3.61 s ± 55.9 ms. For a large deployment of the proposed solution where reducing computational cost was important, the Checksum method could be used first followed by the Luminance Difference Method only on the image sets that changed for the Checksum method. 

These errors show that the metrics used to determine if a data link is from a Network Camera can improve. Potential solutions to these problems are discussed in Section~\ref{sec:future_work}.

In the future, more frames can be taken to further improve the accuracy of the Network Camera identification. Next we will discuss a few sample websites from the crawl.

\subsection{Website Result Analysis}
\label{subsec:website_examples}

In this section, we review examples from two categories of the 73 tested domains. In \ref{sssec:few_cameras} we will discuss sites where the crawler does not find any camera data. In \ref{sssec:dup_cameras} we will discuss the sites where the Camera Discovery module collects significantly more data then expected, given the manual estimate. In these examples, duplicate  data links are found on the site. 

\subsubsection{Few Cameras Found}
\label{sssec:few_cameras}

No camera data is found on 26 of the 73 sites studied. For 9 of these sites, the Web Crawler is unable to identify any camera data or streaming links on any pages. In many cases, this is because the site did not properly load in the browser, or the website would exceed the 180 second timeout of the crawler. 

In a few cases, the crawler is able to identify the camera data but the data does not change during the crawl. This is the case for the Indiana Department of Transportation website \cite{pws_trafficwise_org}. During the crawl, the links to the camera data on the site are discovered by the crawler. No cameras are identified from the site because the cameras stop updating before the frames are taken.  The site later went offline for maintenance.

On The Snow~\cite{onthesnow}, a travel camera website, is another example of a site where no cameras are found even though 5,743 pages are crawled and 20,264 images are discovered. Many of the cameras are not directly embedded in the site and the user must click on a play button where a time-lapse of images from the last 24 hours are shown in a short video. 

Another site that the crawler is unable to collect all the camera data is the WorldCam website~\cite{worldcam_eu}. This site has 14,847 cameras listed, many of which come from the community posting links to the cameras. Almost none of the cameras on this site are embedded directly on the site. Only 164 cameras are found on this site. This is because the crawler is limited to following links only from the initial domain. 

In three websites, the method used to distribute the Network Camera data is different then those discussed in Section~\ref{subsec:camera_formats}. On these three sites the format of the URL is similar to URL format (\ref{url:queryandtimestamp}) from Section~\ref{subsec:camera_formats} however, an HTTP request sent $t_0+\Delta t$ will return the same data as a request sent at time $t_0$. These sites follow the URL format shown in (\ref{url:iddatetime}). Where the \verb|<date-time>| field is updated each time a new image is posted. For example, on Montana Department of Transportation's website ~\cite{rwis_mdt_mt_gov}, to get data from November, 4th, 2018 at 6:08 PM, the \verb|<date-time>| field of the URL in~(\ref{url:iddatetime}) must be set to \verb|2018-11-04_18-08-00.jpg|. 

\begin{urlverb}
    <base URL>/<id>_<date-time>.jpg \label{url:iddatetime}
    <base URL>?<id>_<date-time>.jpg \label{url:idplusquery}
\end{urlverb}

Another example can be seen on oktraffic.org \cite{oktraffic}. The format of the data links on this site can be seen in~(\ref{url:idplusquery}). Here, a date-time is incremented in the query string for each new image. Changes like these are imperceptible to the user because a JavaScript function automatically generates a new data link each second and updates the image on the page. The format of examples (\ref{url:iddatetime}) and (\ref{url:idplusquery}) does not follow our definition of a Network Camera because new HTTP requests to these data links will not yield new image data. The only way to tell if a data link behaves like (\ref{url:queryandtimestamp}) or like (\ref{url:idplusquery}) is to send an HTTP request with and without the query string and see what kind of data is returned. This behavior in these examples does not follow our definition of a Network Camera from Section~\ref{sec:introduction} so these Network Cameras will not be found by our system. 

These examples showcase the most prevalent errors with the proof-of-concept experiment. In some cases, finding the cameras on these websites relies on giving the Web Crawler permission to visit additional domains. In other cases, the crawler will need to be outfitted with additional methods of extracting and identifying camera data. 

\subsubsection{Duplicate Cameras Found}
\label{sssec:dup_cameras}

On 23 sites, the number of cameras aggregated by our system was greater than the expected number of cameras on that site. This usually happens because the camera data is assigned multiple different links across the site. The most common example of this are sites that have query strings in the image links, like those discussed in Section~\ref{sec:website_analysis}. For example, \cite{rlp_traffic_cameras, wsdot,hb_511_nebraska,gnb_ca} all have a random number or time-stamp after the camera link.

\begin{figure}[t]
    \centering
    \includegraphics[width=0.75\linewidth]{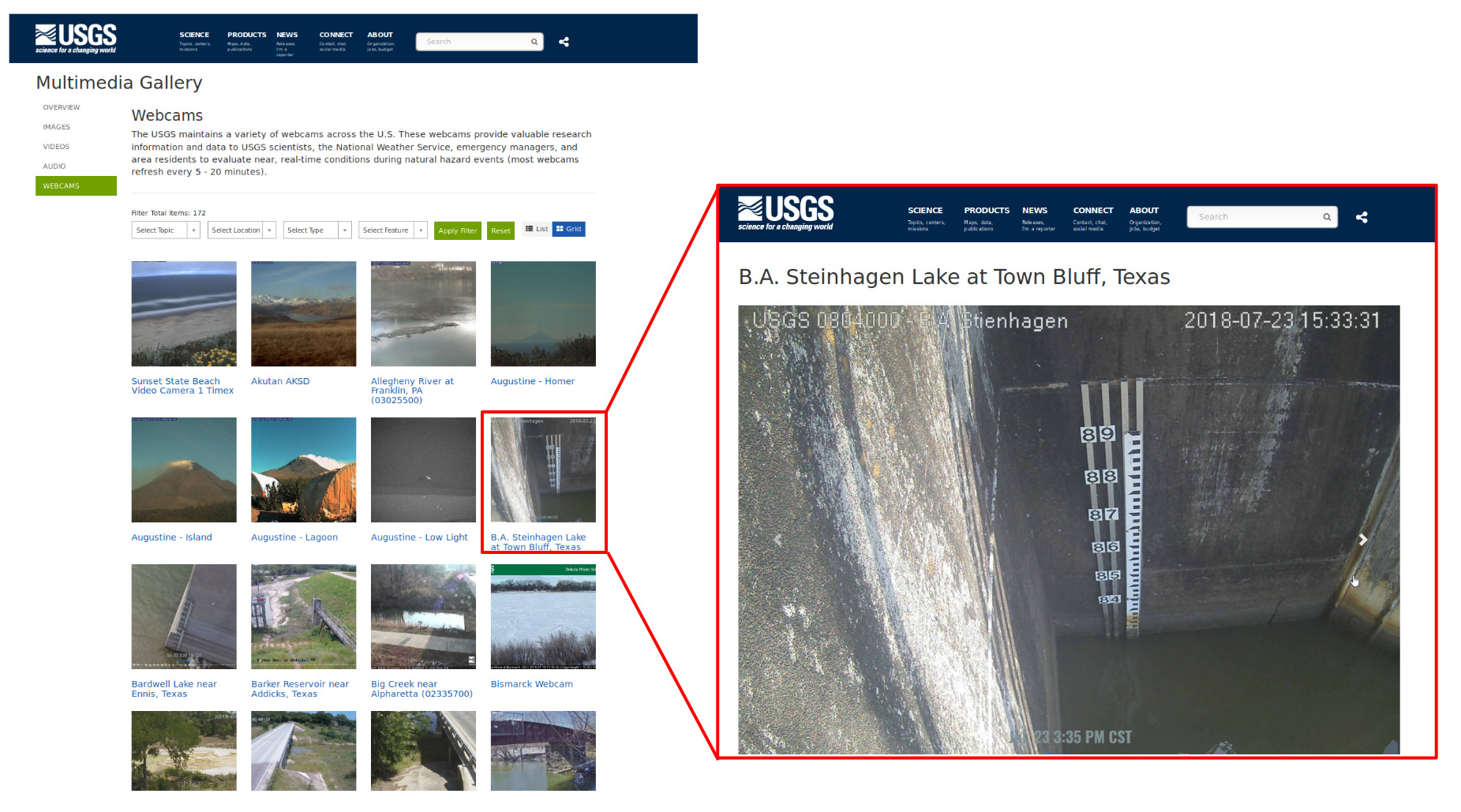}
    \caption{Example of thumbnail images in a list are shown in on the left. The image on the right shows the actual feed of the camera. The thumbnail image has a different data link then the full size image ~\cite{hvo_wr_usgs}.}
    \label{fig:usgs}
\end{figure}

Figure~\ref{fig:usgs} shows an example of thumbnails, another common aspect that causes errors in the Camera Discovery module. Sites with thumbnails ~\cite{hvo_wr_usgs,swisswebcams,kandrive,rlp_traffic_cameras} may have multiple links to the same data. Some are re-sized for display in a list that links to the full size image. 
Some sites, 511 South Carolina site~\cite{511sc} and California Department of Transportation Site ~\cite{dot_ca_gov}, have links to both streaming and static camera data or cameras that have streams available in multiple formats. The Camera Discovery module will aggregate a new camera for every data link found. 

Another common reason that the Camera Discovery module may have found too many cameras on a site is that links to the previous static frames are also found on the site. For example, California Department of Transportation \cite{dot_ca_gov} has links to the past 12 updates listed on the site. The Camera Discovery module counted a new camera for each of these links causing an overestimate of the cameras found. 

The root of this problem is that the Camera Discovery module can only tell if two potential camera links are the same by looking at the URL to the image data. The current method offers no way to tell if a camera has been found before on the same site or a different site. Possible solutions to this problem are discussed in Section~\ref{sec:future_work}.

\section{Future Work}
\label{sec:future_work}

In future work, we plan to implement several improvements into the Network Camera Discovery System presented in this work. Improvements could be made to allow the crawler to achieve a higher accuracy of camera classification. The current method of comparing image luminance is simple and a more sophisticated algorithm could be created that would reduce false-positives like those in Figure~\ref{fig:error_examples}. This problem could also be solved by taking more frames from the data links or by training a machine learning model to recognize Network Camera images. 
The current Camera Discovery module has no way of checking if two data links point to the same camera data. This problem is difficult to solve as some cameras are not static and move their viewpoint over time.

\section{Conclusion}
\label{sec:conclusion}

In conclusion, this work provides an in-depth look at how Network Camera information is distributed on the Internet. We provide a method for aggregating this valuable data source automatically and methods to determine on a given web page, what data on the page is from a Network Camera. We present a proof-of-concept version of the Network Camera discovery system that was successfully able to identify 57,364 Network Camera data links from 237,257 unique web pages over a 55 day test run. 
The data found by the Automated Network Camera Discovery system presented in this work can create a central repository of the thousands of public Network Cameras all around the world. This data can be beneficial for a variety of applications that need real-time data.

\bibliographystyle{ACM-Reference-Format}
\bibliography{main,helps_papers,cam_sites}

\end{document}